\documentclass[pra,floatfix,twocolumn,superscriptaddress]{revtex4}

\usepackage{amsmath,amssymb,amsfonts,amsthm}
\usepackage{graphicx} 
\usepackage{dcolumn} 
\usepackage{bm} 
\usepackage[dvipsnames]{xcolor}
\usepackage{bbm}
\usepackage{exscale}
\usepackage{longtable}

\newcommand{\beq}{\begin{equation}} 
\newcommand{\eeq}{\end{equation}}
\newcommand{\bqa}{\begin{eqnarray}} 
\newcommand{\eqa}{\end{eqnarray}}
\newcommand{\nn}{\nonumber}

\newcommand{\dg}{^\dagger}
\newcommand{\rt}[1]{\sqrt{#1}\,}
\newcommand{\smallfrac}[2]{\mbox{$\frac{#1}{#2}$}}
\newcommand{\half}{\smallfrac{1}{2}}

\newcommand{\an}[1]{\langle{#1}\rangle}
\newcommand{\ban}[1]{\big\langle{#1}\big\rangle}

\newcommand{\vdp}{van~der~Pol}

\newcommand{\Tr}{{\rm Tr}}

\newcommand{\ann}{\hat{a}}
\newcommand{\adg}{\hat{a}\dg}

\newcommand{\Hhat}{\hat{H}}

\newcommand{\Vhat}{\hat{V}}

\newcommand{\Dcal}{{\cal D}}
\newcommand{\Lcal}{{\cal L}}

\newcommand{\wo}{\omega_0}

\newcommand{\xhat}{\hat{x}}

\newcommand{\kup}{\kappa_{\uparrow}}
\newcommand{\kdn}{\kappa_{\downarrow}}

\newcommand{\kddn}{\kappa_\Downarrow}

\newcommand{\qhat}{\hat{q}}
\newcommand{\phat}{\hat{p}}
\newcommand{\gammabar}{\bar{\gamma}}
\newcommand{\sqr}{\text{\tiny $\square$}}
\newcommand{\smcirc}{\text{\large $\circ$}}

\newcommand{\dia}{\diamondsuit}
\newcommand{\too}{\longrightarrow}

\begin{document}

\title{Relaxation oscillations and frequency entrainment in quantum mechanics}

\author{A. Chia}
\affiliation{Centre for Quantum Technologies, National University of Singapore}

\author{L. C. Kwek}
\affiliation{Centre for Quantum Technologies, National University of Singapore}
\affiliation{National Institute of Education, Nanyang Technological University, Singapore}

\author{C. Noh}
\affiliation{Kyungpook National University, Daegu, South Korea}

\date{\today}

\begin{abstract}

Frequency entrainment of continuous-variable oscillators has to date been restrained to the weakly-nonlinear regime. Here we overcome this bottleneck and extend frequency entrainment of quantum continuous-variable oscillators to arbitrary nonlinearities. The previously known steady state of such quantum oscillators in the weakly-nonlinear regime (also known as a Stuart--Landau oscillator) is shown to emerge as a special case. Most importantly, the hallmark of strong nonlinearity---relaxation oscillations---is shown in quantum mechanics. Depending on the oscillator's nonlinearity, relaxation oscillations are found to occur via two distinct mechanisms in phase space.

\end{abstract}


\maketitle

\section{Introduction}
\label{Intro}

An exciting and long-standing endeavour in physics is the quest of quantum-mechanical phenomena with classical counterparts. Classical nonlinear systems have been a rich source of inspiration \cite{Str15}. A prominent example of this is chaos \cite{Haa00,Wim14,BHJ00,SM01,EHC17}. More recently, another fascinating aspect of nonlinear dynamics called synchronisation has garnered considerable interest in quantum systems \cite{RB18,LCW14,WWBT17,LANB16,SHM+18,AEAM+15}. In rather broad terms, synch g ronisation studies how oscillating systems modify their rhythms under weak coupling to one another \cite{PRK01}. Here we consider the simplest such scenario where a quantum oscillator is driven classically and investigate frequency entrainment \cite{BJPS09,PRK01}---the modification of the oscillator's frequency under driving \footnote{That is, we do not consider the phase dynamics of the oscillator and issues related to phase locking. Some may argue that a description of synchronisation should involve discussions of both frequency entrainment and phase locking, the latter of which we do not provide.}. Classical synchronisation theory provides a unifying view of seemingly different processes, from the unison flashing of fireflies to the adjustment of circadian rhythms \cite{BJPS09,PRK01}.

An important criterion for oscillators to be fit for synchronisation is that it must possess a limit cycle i.e.~it exhibits self-sustained oscillations \cite{AVK66}. An archetypal example of such a system is the Stuart--Landau oscillator \cite{Lan44,Stu60,Kur84}, defined by the following differential equation for its complex amplitude $\alpha$,
\begin{align}
\label{CSL}
	\alpha' = - i \, \wo \, \alpha + \frac{\gamma_1}{2} \, \alpha - \gamma_2 \,  |\alpha|^2 \, \alpha  \; ,
\end{align}
where $\wo$ is its frequency of oscillation and we require both $\gamma_1 > 0$ and $\gamma_2 > 0$. We are using a prime to denote differentiation with respect to time, i.e.~$\alpha'=d\alpha/dt$. The extension of synchronisation to quantum systems is then made possible on quantising \eqref{CSL}. This requires the existence of a physically valid generator of time evolution such that its dynamics follows \eqref{CSL} on average. It is straightforward to show that such evolution for the oscillator's state $\rho$ can be obtained by adding one-photon gain and two-photon loss to a harmonic oscillator of frequency $\wo$ \cite{LS13,WNB14},
\begin{align}
\label{QSL}
	\rho' = \Lcal_\smcirc \rho = -i \, \wo \, \big[\adg\ann,\rho\big] + \gamma_1 \, \Dcal\big[\adg\big] \rho + \gamma_2 \, \Dcal\big[\ann^2\big] \rho  \; ,
\end{align}
where $[\ann,\adg]=\hat{1}$. We have defined the shorthand $\Dcal[\hat{c}]$ for any $\hat{c}$ to be
\begin{align}
\label{D[c]}
	\Dcal[\hat{c}] \rho = \hat{c} \, \rho \, \hat{c}\dg - \frac{1}{2} \; \hat{c}\dg \hat{c}\, \rho - \frac{1}{2} \; \rho \, \hat{c}\dg \hat{c}  \; .
\end{align}

Equation \eqref{QSL} has enjoyed widespread success in quantum synchronisation and related studies due to its simple form while capturing a sufficient amount of nontrivial physics, see e.g.~\cite{LS13,WNB14,RB18,LCW14,WWBT17,LANB16,SHM+18,WNB15,IK17,DC19,KN20}. Despite this, \eqref{QSL} is not without drawbacks. A major limitation of \eqref{QSL} is that it is only valid for weakly-nonlinear oscillators. This restriction on \eqref{QSL} follows from a well-known result in nonlinear dynamics where the weakly-nonlinear limit of a class of limit-cycle oscillators is shown to be \eqref{CSL} \cite{Lan44,Stu60,Kur84}. As part of this paper, we will remove this limitation.

Physically, limit-cycle behaviour can only occur if a system has energy gain to sustain the oscillations. It must also have a mechanism for energy loss to balance the energy gain so that stable oscillations can be achieved. Limit-cycle oscillators are thus nonconservative systems. One way to model the interplay between gain and loss in such nonconservative systems is to generalise a damped harmonic oscillator to allow for an arbitrary nonlinear friction term $\gamma(x,x')$:
\begin{align}
\label{2ndOrder}
	x'' + \wo^2 \, x + \mu \, \gamma(x,x') = 0  \; .
\end{align}
Here $\mu$ is called the nonlinearity parameter. The weakly-nonlinear limit is then defined by $0<\mu\ll1$ and it is in this regime that \eqref{CSL} is confined to \footnote{This is a nontrivial result from the analysis of bifurcations and its corresponding normal form where $\mu$ is the bifurcation parameter. The result is that the Stuart--Landau equation is the normal form for a Hopf bifurcation at $\mu=0$. Two types of Hopf bifurcation may occur, supercritical and subcritical. In a supercritical Hopf bifurcation at $\mu=0$, a stable fixed point at the origin of phase space becomes unstable while a stable limit cycle around it is simultaneously formed. It is actually this special case of a supercritical Hopf bifurcation that \eqref{CSL} corresponds to.}. Examples of $\gamma(x,x')$ that lead to self-sustained oscillations are the Rayleigh and  \vdp\ oscillators, both of which will be studied in this paper.

Note the existing quantum synchronisation literature refers to \eqref{CSL} and \eqref{QSL} as a classical and quantum \vdp\ model respectively, such as Refs.~\cite{LS13,WNB14,RB18,LCW14,WWBT17,LANB16,SHM+18,WNB15,IK17,DC19,KN20}. This is a misnomer because the Stuart--Landau model does not correspond uniquely to the \vdp\ oscillator for reasons just explained. Since we will be studying the exact \vdp\ oscillator, this distinction between the Stuart--Landau and the \vdp\ models should be kept in mind to avoid any confusion with the previous literature.

Despite the growing interest in quantum synchronisation and the general importance of limit-cycle oscillators, the extension of these concepts to quantum mechanics for arbitrary nonlinearities has remained elusive. This bottleneck arises because the nonlinear friction term in \eqref{2ndOrder} makes its quantisation difficult. Thus, several basic but important questions remain unanswered. The first is what properties can we expect of the quantum limit cycle once we leave the regime of vanishing nonlinearity? Although the corresponding classical limit cycles might give us a hint, it is not at all clear how they would translate to the quantum realm \cite{SCVC15}. The smoking gun that one has accessed the strongly-nonlinear regime is relaxation oscillations. These are oscillations where two widely separated timescales coexist. The heartbeat is a physical example \cite{vdPvdM28,Cro77}. Furthermore, assuming one can quantise a general nonlinear oscillator with arbitrary nonlinearity, how would relaxation oscillations be contained in such a model? What does frequency entrainment look like in phase space? In this paper we hope to shed some light on these issues.

A preview of our paper is as follows. The Rayleigh and \vdp\ oscillators are introduced in Sec.~\ref{Class_Ray+vdP}. Here we explain their phase-space dynamics focusing on their limit-cycle behaviour and relaxation oscillations as these aspects will feature prominently later in their quantum analogues. These models are then quantised in Sec.~\ref{Quant_Ray+vdP}. The physics produced by the quantum models are left to Sec.~\ref{PhysResults} where their steady states for different strengths of the nonlinearity are considered. The sense in which they correspond to their classical counterparts will also be discussed. Finally in Sec.~\ref{Qsync} we apply our model to extract relaxation oscillations and frequency entrainment. Here we simply add an external drive to the Rayleigh oscillator and look at how much the oscillator's frequency has changed for varying drive strengths. We have kept our exposition of frequency entrainment to an elementary level here because the numerics in the strongly-nonlinear regime is much more intensive than the weakly-nonlinear regime. Also, we do not discuss phase locking. It is for this reason that we refer to our results for a driven oscillator as frequency entrainment rather than synchronisation where phase locking may play some role. We then show that for a strongly-nonlinear oscillator, its dynamics under driving exhibit relaxation oscillations. Interestingly, these relaxation oscillations are realised by a sort of slimy behaviour in phase space and occur in two distinct modes depending on its nonlinearity relative to the drive. We conclude in Sec.~\ref{Discussion} with a summary and mention some forthcoming work.

\section{Classical models}
\label{Class_Ray+vdP}

\subsection{Rayleigh oscillator}
\label{RayClass}

\begin{figure}[t]
\centerline{\includegraphics[width=0.49\textwidth]{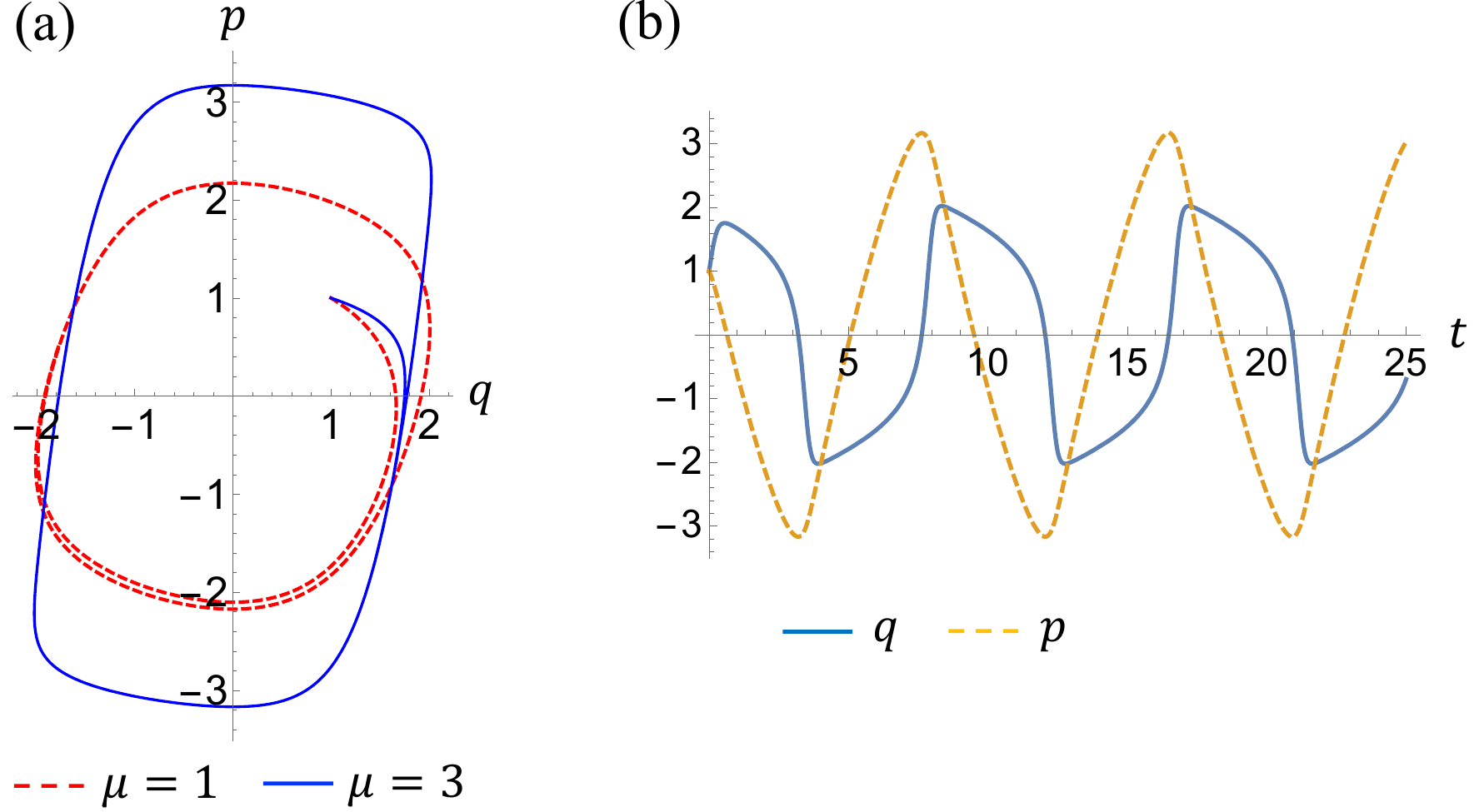}}
\caption{\label{RayLC} Dynamics of the Rayleigh oscillator for $q_0=\wo=1$ and $q(0)=p(0)=1$. (a) Limit cycles of the Rayleigh oscillator for $\mu=1$ (red dashed) and $\mu=3$ (blue solid). (b) Evolution of $q$ (blue solid) and $p$ (yellow dashed) corresponding to the $\mu=3$ limit cycle in (a). }
\end{figure}

A model of a limit-cycle oscillator was given by Lord Rayleigh in as early as 1883 \cite{Ray1883}, who also emphasised the analogy between sound waves and electrical oscillations \cite{Lan96,Ray45}. Rayleigh considered a nonlinear friction term in \eqref{2ndOrder} given by \cite{JS07}
\begin{align}
\label{RayleighDefn}
	 \gamma(x,x') = a \, \frac{x'^{\,3}}{3} - b \, x'  \; ,
\end{align}
where $a>0$ and $b>0$. We shall work with dimensionless position and momentum variables $q$ and $p$ respectively. It will also be convenient to recast the second-order equation \eqref{2ndOrder} as two coupled first-order equations. In this case the Rayleigh oscillator is defined by the evolution of $p$ in
\begin{align}
\label{RectModelDefn}
	q' = \wo \, p  - \mu \, f(q)  \; ,   \quad   p' = - \wo\,q   \; .
\end{align}
Here we have defined the  
\begin{align}
	f(q) = \frac{q^3}{3} - q^2_0 \, q \; . 
\end{align}
This corresponds to \eqref{RayleighDefn} with $a=1/\wo^2$ and $b=q^2_0$. The limit cycles of the Rayleigh oscillator are shown in Fig.~\ref{RayLC}\,(a) for $\mu=1$ and $\mu=3$. We then plot, for the $\mu=3$ case the evolution of the dimensionless position and momentum in Fig.~\ref{RayLC}\,(b). A very short transient can be observed in Fig.~\ref{RayLC}\,(b) since we did not start on the limit cycle. From looking at both Fig.~\ref{RayLC}\,(a) and (b) it is clear that the Rayleigh oscillator spends most of its time along the two vertical sides of its limit cycle for large $\mu$. Moreover, relaxation oscillations can now be seen explicitly in the position of the Rayleigh oscillator where a separation of timescales is clearly evident from Fig.~\ref{RayLC}\,(b). Note also that as we increase $\mu$, the limit cycle tilts slightly towards the left (for fixed $q_0$).
\begin{figure}[t]
\centerline{\includegraphics[width=0.38\textwidth]{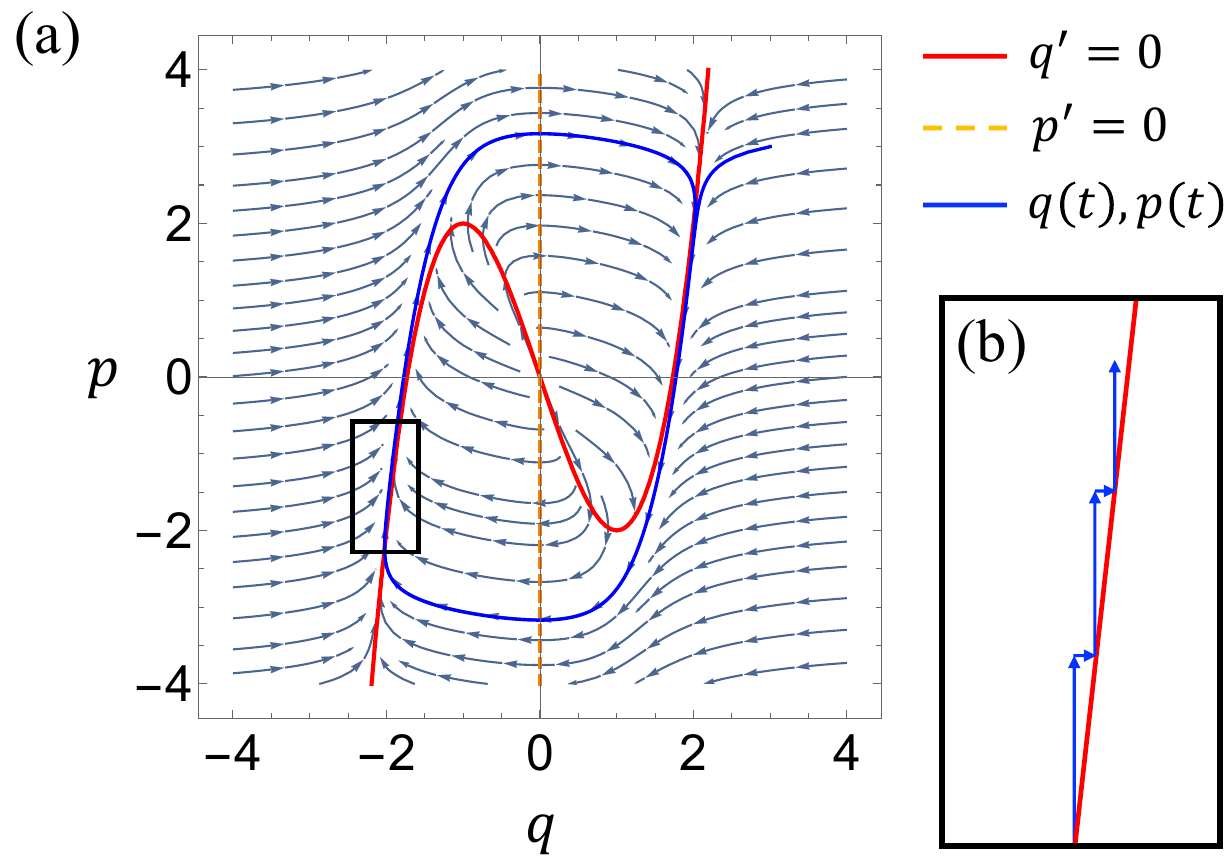}}
\caption{\label{RayFlow} Flow and nullclines for the Rayleigh oscillator corresponding to $q_0=\wo=1$, and $\mu=3$, starting at $q(0)=p(0)=3$. (a) Flow defined by $(q',p')$, the $q$ nullcline (red solid curve), $p$ nullcline (yellow dashed curve, at exactly $q=0$), and the limit cycle (blue solid curve). The $q$ nullcline has turning points at $\big(\!-\!q_0,\mu q^2_0(q_0-3)/3\wo\big)$ (second quadrant) and $\big(q_0,-\mu q^2_0(q_0-3)/3\wo\big)$ (fourth quadrant). (b) Magnified view of the flow near the $q$ nullcline within the black box in (a). The staircase movement of the phase-space point is only an explanation of its motion near the $q$ nullcline, not the true phase-space trajectory which is a smooth curve.}
\end{figure}

To see why the Rayleigh oscillator spends significantly more time along the two vertical sides of its limit cycle it is helpful to consider the flow and nullclines defined by the right-hand sides of \eqref{RectModelDefn}. In Fig.~\ref{RayFlow}\,(a) we plot the $q$ nullcline (red solid line, defined by $q'=0$), and the $p$ nullcline (yellow dashed line, defined by $p'=0$) corresponding to Fig.~\ref{RayLC}\,(b) together with the flow [defined by $(q',p')$] and its associated limit cycle (blue solid curve). In Fig.~\ref{RayFlow}\,(a) we see the flow lines point towards the two vertical sides of the $q$ nullcline. In Fig.~\ref{RayFlow}\,(b) we show how the phase-space trajectory travels along the $q$ nullcline [corresponding to the region of phase space inside the black box in Fig.~\ref{RayFlow}\,(a)]: When the phase-space point reaches the $q$ nullcline, the only way it can move is to go directly up in the positive $p$ direction. However, when the phase-space point moves upwards it experiences a push towards the right (positive $q$ direction). This pushes the phase-space point back onto the $q$ nullcline where the only possible movement is again in the positive $p$ direction. This produces the staircase motion along the $q$ nullcline shown in Fig.~\ref{RayFlow}\,(b). Of course, in reality the phase-space point moves in continuous time and the trajectory is a smooth curve that hugs the $q$ nullcline. We can now see how the slow and fast timescales are produced in the Rayleigh oscillator. The phase-space point makes this ``staircase'' climb along the $q$ nullcline (slow timescale) until it reaches its turning point where the $q$ nullcline ends and it is free to move horizontally (fast timescale). From this we can also understand the elongation of the limit cycle when $\mu$ is increased. For a fixed $q_0$, the two turning points of the $q$ nullcline is shifted farther along $p$ (one in the positive direction and one in the negative direction), and the phase-space point is blocked from moving horizontally for a greater range of $p$ values.

The $q$ nullcline also explains why the limit cycle tilts towards the left when we increase $\mu$. For large $\mu/\wo$ and fixed $q_0$, the slope of the two elongated sides of the limit cycle is dictated by the slope of the two vertical sides of the $q$ nullcline [see Fig.~\ref{RayFlow}\,(a)].
The slope of the vertical branches of the $q$ nullcline in Fig.~\ref{RayFlow} becomes steeper and steeper as $\mu/\wo$ becomes larger and larger. This in turn causes the limit cycle to become more and more vertical which makes it tilt more and more to the left.

\subsection{\vdp\ oscillator}
\label{ClassvdP}

Historically the \vdp\ oscillator was proposed to describe electrical oscillations seen in triode circuits in the 1920s  \cite{vdP20,Car60}. Arguably, it has since become the most well-known example of a self-sustained oscillator, and its name has become synonymous with relaxation oscillations \cite{vdP26,GL12}. The model corresponds to a nonlinear damping term in \eqref{2ndOrder} of the form,
\begin{align}
	\gamma(x,x') = \big( x^2 - q^2_0 \big) \, x' \; . 
\end{align}
Note that \eqref{2ndOrder} with this form of $\gamma(x,x')$ can be obtained by differentiating the second-order equation for the Rayleigh oscillator again and then letting $x'\too x$ and $a=1$. In modern applications, the \vdp\ and Rayleigh oscillators are often fused together to describe human and robotic limb movements \cite{Fuc13}. Such Rayleigh--\vdp\ oscillators are also widely used in biological cybernetics to model interpersonal synchronisation \cite{KRK19} (see Ref.~\cite{ABB16} and the references therein). 

\begin{figure}[t]
\centerline{\includegraphics[width=0.49\textwidth]{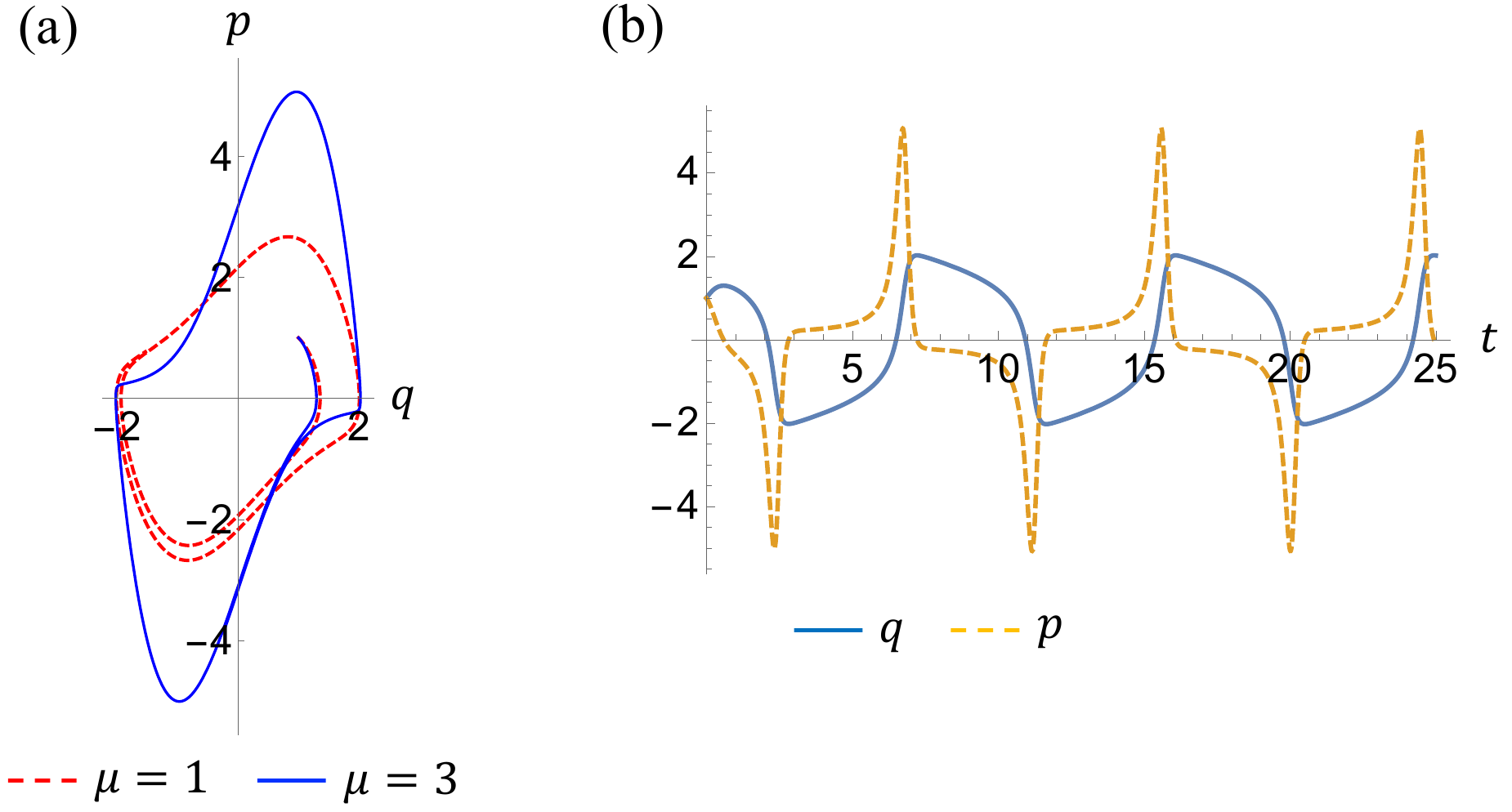}}
\caption{\label{vdPLC} Dynamics of the \vdp\ oscillator for $q_0=\wo=1$ and $q(0)=p(0)=1$. (a) Limit cycles of the \vdp\ oscillator for $\mu=1$ (red dashed) and $\mu=3$ (blue solid). (b) Evolution of $q$ (blue solid) and $p$ (yellow dashed) corresponding to the $\mu=3$ limit cycle in (a).}
\end{figure}
Again, it is convenient to recast the \vdp\ oscillator as a set of first-order equations,
\begin{align}
\label{DiamondModelDefn}
	q' = \wo \, p \; ,  \quad  p' = -\mu\,\gamma(q,p) - \wo \, q  \;.
\end{align}
Note that we have written \eqref{RectModelDefn} and \eqref{DiamondModelDefn} so that they share the same second-order differential equation for $q$, but have different second-order equations for $p$. In general an $n$-dimensional nonlinear system is defined by a set of $n$ first-order differential equations rather than a single $n$th-order equation \footnote{In general, a given $n$th-order differential equation does not correspond uniquely to a set of $n$ coupled first-order equations. Furthermore, a given set of coupled first-order equations cannot always be decoupled to give a single second-order equation. Thus the definition of a general nonlinear system should always be in terms of a set of first-order equations.}.

\begin{figure}[t]
\centerline{\includegraphics[width=0.38\textwidth]{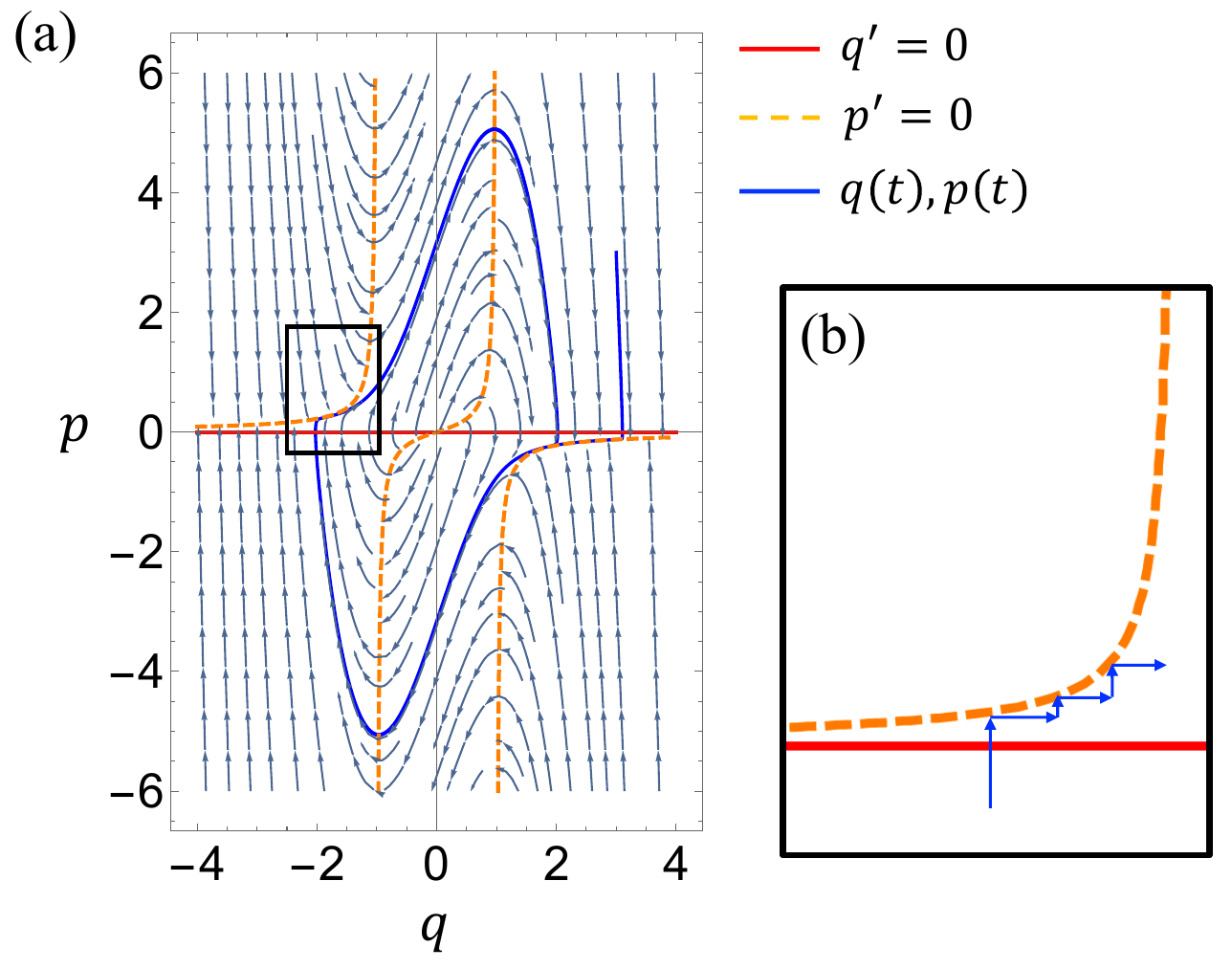}}
\caption{\label{vdPFlow} Flow and nullclines for the \vdp\ oscillator corresponding to $q_0=\wo=1$, and $\mu=3$, starting at $q(0)=p(0)=3$. (a) Flow given by $(q',p')$, the $q$ nullcline (red solid curve), $p$ nullcline (yellow dashed curve, at exactly $q=0$), and the limit cycle (blue solid curve). (b) Magnified view of the flow near the $q$ and $p$ nullclines within the black box in (a). The staircase movement of the phase-space point is only an explanation of its motion near the $p$ nullcline, not the true phase-space trajectory which is a smooth curve.}
\end{figure}
The limit cycle of the \vdp\ oscillator for $\mu=1$ and $\mu=3$ are shown in Fig.~\ref{vdPLC}\,(a) and the evolution of its dimensionless position and momentum for $\mu=3$ are displayed in Fig.~\ref{vdPLC}\,(b). Note that in contrast to the Rayleigh oscillator, relaxation waveforms can be seen in both $q$ and $p$ in the \vdp\ oscillator. Looking at Fig.~\ref{vdPLC}\,(a) and (b) together we find the slow sections of the limit cycle to be in the second and fourth quadrants of phase space where the limit cycle bends inwards near the $q$ axis. As before, looking at its nullclines and flow pattern provide useful insight on its limit-cycle behaviour. We show this in Fig.~\ref{vdPFlow}\,(a) along with a magnified view of the slow portion of the limit cycle in Fig.~\ref{vdPFlow}\,(b). We observe in Fig.~\ref{vdPFlow}\,(b) how the $q$ nullcline (red solid line) admits flow only in the positive $p$ direction. But shortly after crossing the $q$ nullcline the upwards movement is blocked by the $p$ nullcline (yellow dashed curve). In a similar fashion as the Rayleigh oscillator in Fig.~\ref{RayFlow}\,(b), the $p$ nullcline and the upwards flow underneath it creates a slow timescale. We then find that a fast timescale sets in when the $p$ nullcline bends upward and no longer blocks the flow going upwards.

Lastly, an important feature of limit-cycle oscillators to note is that their frequencies can be modified in the presence of nonlinearity \cite{Str15,Sco05}. This is an intrinsic property of the oscillator itself, not related to frequency entrainment where an external force is applied. It will therefore be interesting to see if such frequency shifts may also be observed in quantum mechanics.

\section{Quantum models}
\label{Quant_Ray+vdP}

\subsection{Rayleigh oscillator}

We consider the quantisation of the Rayleigh oscillator first. In a quantum model we have to treat $q$ and $p$ as operators and seek a generator of time evolution such that \eqref{RectModelDefn} is reproduced on average. By definition,
\begin{align}
\label{1stOrderRectModel}
	\an{\qhat}' = \wo \, \an{\phat}  - \mu \, \an{f(\qhat)}  \;,   \quad
	\an{\phat}' = - \wo \, \an{\qhat}   \;.
\end{align}
It will be convenient to define 
\begin{align}
	\ann = \frac{1}{2} \; ( \qhat + i \phat)  \; .
\end{align}
Note that on using $[\ann,\adg]=\hat{1}$, the commutator for our dimensionless position and momentum becomes 
\begin{align}
	[\qhat,\phat] = 2 i \hat{1} \; . 
\end{align}
Equation \eqref{1stOrderRectModel} can then be compactly written as a single equation using the annihilation operator,
\begin{align}
\label{adotRectModel}
	\an{\ann}' = {}& - i\,\wo\, \an{\ann} + \frac{\mu}{2} \, ( q^2_0 - 1 ) \, \big[ \an{\ann} + \an{\adg} \big]  \nn \\
	                             & - \frac{\mu}{6} \, \big[ \an{\ann^3} + \an{\adg{}^3} \big]  
	                                 - \frac{\mu}{2} \, \big[ \an{\adg \, \ann^2} + \an{\adg{}^2 \, \ann} \big]  \;.
\end{align}
This is now the defining equation of motion for the quantum Rayleigh oscillator. The task at hand is then to find a $\rho'$ such that $\an{\ann}'=\Tr[\ann\rho']$ has the form of \eqref{adotRectModel}. This has been a major stumbling block in quantum synchronisation and nonlinear dynamics. A generator of time evolution that correctly produces \eqref{adotRectModel} is given by the following equation 
\begin{align}
\label{Lsqr}
	\rho' = \Lcal_\sqr \, \rho = {}& - i \, \big[ \Hhat_\sqr\,, \rho \big] + \mu\,\big(q^2_0-1\big) \, \Dcal\big[ \adg \big] \rho   \nn \\
	                                                       & + \frac{3\mu}{4} \; \Dcal\big[\ann^2\big] \rho + \mu \; \Dcal\big[\adg\ann-\half\,\adg{}^2\big] \rho   \; ,
\end{align}
where $\Dcal[\hat{c}]$ is as defined in \eqref{D[c]}. We have used a square subscript to remind us that its classical limit cycle has the shape of ``square'' for large $\mu$. Here $\Hhat_\sqr$ is Hermitian, given by
\begin{align}
\label{Hsqr}
	\hat{H}_\sqr = {}& \wo \, \adg\ann + i\,\zeta_\sqr \, \big( \adg \ann^3 - \adg{}^3 \ann \big) + i\,\beta_\sqr \, \big( \ann^4 - \adg{}^4 \big)  \nn \\
	                 & - i \, \eta_\sqr \, \big( \ann^2 - \adg{}^2 \big)  \;,
\end{align}
with 
\begin{align}
\label{RayCoeff}
	\zeta_\sqr = \frac{\mu}{12}  \; ,  \quad \beta_\sqr = \frac{\mu}{24}  \; ,  \quad  \eta_\sqr = \frac{\mu\,(q^2_0-1)}{4} \; .
\end{align}
It is straightfoward to show that \eqref{Lsqr}--\eqref{RayCoeff} satisfies \eqref{adotRectModel}. From the definition of $\an{\ann}'$\, and using the proposed $\Lcal_\sqr$\,,
\begin{align}
\label{<a'>Ray}
	\ban{\ann}' = {}& - i \, \Tr\Big\{ \ann \, \big[ \Hhat_\sqr,\rho\big] \Big\} +  \mu \, \Tr\Big\{ \ann \, \Dcal\big[\, \adg\ann - \half\,\adg{}^2 \,\big] \rho \Big\}  \nn \\
	                               & + \mu \, \big( q^2_0 - 1 \big)  \, \Tr\Big\{ \ann \, \Dcal\big[ \adg \big] \rho \Big\} + \frac{3 \mu}{4} \, \Tr\Big\{ \ann \, \Dcal\big[ \ann^2 \big] \rho \Big\}  \: .
\end{align}
Using \eqref{Hsqr} we find the Hamiltonian contribution to be
\begin{align}
\label{Ham1Ray}
	-i \, \Tr\Big\{ \ann \, \big[ \Hhat_\sqr,\rho\big] \Big\} 
	= {}& - i \, \wo \Tr\Big\{ \ann \, [\adg \ann, \rho ] \Big\}  \nn \\
	       & - \, \eta_\sqr  \, \Tr\Big\{ \ann \, \big[ \ann^2 - \adg{}^2, \rho\big] \Big\}  \nn \\
	       & +  \beta_\sqr \, \Tr\Big\{ \ann \, \big[ \ann^4 - \adg{}^4 , \rho \big]\Big\} \nn \\
	       & + \zeta_\sqr  \, \Tr\Big\{ \ann \, \big[ \adg\ann^3-\adg{}^3\ann, \rho \big] \Big\}  \, .
\end{align}
It is not difficult to show, term by term,
\begin{align}
\label{Ham2Ray}
	i \, \wo \, \Tr\Big\{ \ann \, \big[\adg \ann, \rho \big] \Big\} = {}&   - i \, \wo \, \ban{\ann}  \; ,  \\
\label{etaTrRay}
	\eta_\sqr  \, \Tr\Big\{ \ann \, \big[ \ann^2 - \adg{}^2, \rho\big] \Big\} = {}&  \frac{\mu}{2} \,  (q^2_0 - 1) \, \ban{\adg}  \; ,  \\
\label{betaTrRay}
	\beta_\sqr \, \Tr\Big\{ \ann \, \big[ \ann^4 - \adg{}^4 , \rho \big]\Big\} = {}& - \frac{\mu}{6} \, \ban{\adg{}^3}  \; ,  \\
\label{zetaTrRay}
	\zeta_\sqr  \, \Tr\Big\{ \ann \, \big[ \adg\ann^3-\adg{}^3\ann, \rho \big] \Big\} = {}& \frac{\mu}{12} \, \ban{\ann^3} - \frac{\mu}{4} \, \ban{\adg{}^2 \ann} \; .
\end{align}
Similarly, the dissipative terms are given by
\begin{align}	
\label{Dis1Ray}
	\mu \, \Tr\Big\{ \ann \, \Dcal\big[\, \adg\ann - \half\,\adg{}^2 \,\big] \rho \Big\} 
	=  {}& \frac{\mu}{4} \, \big[ \, \ban{\adg \ann^2} - \ban{\adg{}^2 \ann} \, \big]  \\
	    {}& - \frac{\mu}{4} \; \ban{\ann^3} \,  \; ,   \nn \\[0.15cm]
	 \mu \, \big( q^2_0 - 1 \big)  \, \Tr\Big\{ \ann \, \Dcal\big[ \adg \big] \rho \Big\} = {}& \mu \, \big( q^2_0 - 1 \big) \, \ban{\ann}  \; ,  \\ 
\label{Dis3Ray} 
	 \frac{3 \mu}{4} \, \Tr\Big\{ \ann \, \Dcal\big[ \ann^2 \big] \rho \Big\} = {}& - \frac{3\mu}{4} \, \ban{\adg \ann^2}  \; .
\end{align}
We then recover \eqref{adotRectModel} on substituting \eqref{Ham1Ray}--\eqref{Dis3Ray} into \eqref{<a'>Ray}.

\subsection{\vdp\ oscillator}

Quantising the \vdp\ model is slightly more complicated due to the nonlinear friction term $\gamma(q,p)$ in \eqref{DiamondModelDefn}. As $\qhat$ and $\phat$ do not commute, the nonlinear friction must be defined with respect to a particular ordering in quantum mechanics. Here we adopt symmetric ordering (also known as Weyl ordering) \cite{Hal13}. The quantum \vdp\ oscillator is therefore defined by 
\begin{align}
	\ban{\qhat}' = \wo \, \ban{\phat} \; ,  \quad  \an{\phat}' = -\mu \, \ban{\bar{\gamma}(\qhat,\phat)} - \wo \, \ban{\qhat} \; ,
\end{align}
where the nonlinear damping coefficient is defined by
\begin{align}
	\gammabar(\qhat,\phat) = \frac{1}{3} \; \big( \qhat^2 \,\phat + \qhat\,\phat\,\qhat + \phat\,\qhat^2 \big) - \mu \, q^2_0 \, \phat \; .
\end{align}
As with the Rayleigh oscillator, we convert these expressions from $\qhat$, $\phat$ to $\ann$ and $\adg$. The amplitude equation of motion for the \vdp\ oscillator is then
\begin{align}
\label{d<a>/dtvdP}
	\ban{\ann}' = & - i \, \wo \, \ban{\ann} + \frac{\mu}{2} \; \big(q^2_0-1\big) \big[ \ban{\ann} - \ban{\adg} \big]  \nn \\
	                           & - \frac{\mu}{2} \; \big[ \ban{\ann^3} - \ban{\adg{}^3} \big] - \frac{\mu}{2} \; \big[ \ban{\adg \, \ann^2} - \ban{\adg{}^2\, \ann} \big]   \; .
\end{align}
The similarity of \eqref{d<a>/dtvdP} to the Rayleigh amplitude equation \eqref{adotRectModel} allows us to quantise the \vdp\ oscillator by making the simple transformation  in $\Lcal_\sqr$:
\begin{align}
\label{Rect<=>Diamond}
	 \big( \zeta_\sqr, \beta_\sqr, \eta_\sqr\big)  \; \longrightarrow \; \big( \zeta_\dia,\beta_\dia,\eta_\dia\big)=\big( -\!3\,\zeta_\sqr, -3\,\beta_\sqr, -\eta_\sqr\big)   \;.
\end{align}
That is, the quantum \vdp\ model is given by
\begin{align}
\label{Ldia}
	\rho' = \Lcal_\dia \, \rho = {}& - i \, \big[ \Hhat_\dia\,, \rho \big] + \mu\,\big(q^2_0-1\big) \, \Dcal\big[ \adg \big] \rho  \nn \\
	                                                       & + \frac{3\mu}{4} \; \Dcal\big[\ann^2\big] \rho + \mu \; \Dcal\big[\adg\ann-\half\,\adg{}^2\big] \rho   \; ,
\end{align}
where we have used a diamond subscript to remind us that its classical limit cycle has the shape of ``diamond'' for large $\mu$. As just described, the only change from the Rayleigh model is in the Hamiltonian 
\begin{align}
\label{Hdia}
	\hat{H}_\dia = {}& \wo \, \adg\ann + i\,\zeta_\dia \, \big( \adg \ann^3 - \adg{}^3 \ann \big) + i\,\beta_\dia \, \big( \ann^4 - \adg{}^4 \big)  \nn \\
	                 & - i \, \eta_\dia \, \big( \ann^2 - \adg{}^2 \big)  \; .
\end{align}
We can show that this is all that is required to obtain $\Lcal_\dia$ by directly calculating the relevant terms from \eqref{etaTrRay}--\eqref{zetaTrRay},
\begin{align}
\label{etaTrvdP}
	\eta_\dia  \, \Tr\Big\{ \ann \, \big[ \ann^2 - \adg{}^2, \rho\big] \Big\} = {}&  - \, \frac{\mu}{2} \,  (q^2_0 - 1) \, \ban{\adg}  \; ,  \\
	\beta_\dia \, \Tr\Big\{ \ann \, \big[ \ann^4 - \adg{}^4 , \rho \big]\Big\} = {}&  \frac{\mu}{2} \, \ban{\adg{}^3}  \; ,  \\
\label{zetaTrvdP}
	\zeta_\dia \, \Tr\Big\{ \ann \, \big[ \adg\ann^3-\adg{}^3\ann, \rho \big] \Big\} = {}& - \, \frac{\mu}{4} \, \ban{\ann^3} + \frac{3\mu}{4} \, \ban{\adg{}^2 \ann} \; .
\end{align}
In the same fashion as the Rayleigh oscillator we can show, on using \eqref{Ham2Ray}, \eqref{Dis1Ray}--\eqref{Dis3Ray}, and \eqref{etaTrvdP}--\eqref{zetaTrvdP} that $\Tr\big\{\ann\Lcal_\dia\rho\big\}$ reproduces \eqref{d<a>/dtvdP}.

\section{Undriven quantum oscillator}
\label{PhysResults}

\subsection{General considerations}
\label{GeneralRemarks}

Some immediate consequences of our models are noteworthy. First, we have been able to put $\rho'$ in the Lindblad form with time-independent coefficients which guarantees the map $\rho(0)\too\rho(t)$ to be completely positive and trace preserving for $q_0 \ge 1$ \cite{Lin76,BP02}. This makes our result simple to simulate using standard software packages (see e.g.~Refs.~\cite{JNN12,JNN13}). In this regard, we note that a Hamiltonian for the classical \vdp\ oscillator has been obtained previously by using an auxiliary oscillator \cite{SCVC15,CSC18}. However, the position and momentum dynamics of the \vdp\ oscillator must be extracted from different oscillators. Furthermore, each oscillator in this two-oscillator system has unbounded phase-space trajectories which makes its quantum mechanics problematic \cite{SCVC15}. Second, $\rho'$ can be seen to contain terms which have unequal numbers of $\ann$ and $\adg$. This is consistent with the fact that relaxation limit cycles are traversed at non-uniform speeds. If $\rho'$ had only terms with equal numbers of $\ann$ and $\adg$ then there would exist a rotating frame at constant angular velocity in which the phase-space point appears stationary, but this is impossible if the limit cycle is traversed at non-uniform speeds. Thirdly, there is no unique quantum model corresponding to a given classical system even in the absence of any ordering ambiguity in $\qhat$ and $\phat$ like the \vdp\ oscillator. This is because quantum nonlinear systems are defined only by their first moments [\eqref{1stOrderRectModel} or equivalently \eqref{adotRectModel}]. This is explicitly demonstrated in \eqref{DTrans} of Appendix~\ref{InterpretRayleigh}. Another example is Ref.~\cite{CHN+19}.

As we have seen, the forms of $\Lcal_\sqr$ or $\Lcal_\dia$ are much more complicated compared to $\Lcal_\smcirc$. Therefore one might wonder how to interpret these models. Essentially all terms except for $\Dcal[\adg\ann-\adg{}^2/2]$ can be interpreted physically as some kind of parametric process. We comment on these in Appendix~\ref{InterpretRayleigh} where we also show how $\Dcal[\adg\ann-\adg{}^2/2]$ can be interpreted as a continuous measurement followed by feedback via squeezing \cite{WM10,Jac14,CW11a,CW11b}. We show in Appendix~\ref{vdPApp} that the same technique with slight modifications permits both $\Dcal[\adg\ann-\adg{}^2/2]$ and $\Hhat=i\,\zeta_\dia \,( \adg \ann^3 - \adg{}^3 \ann)$ to be realised for the \vdp\ oscillator. The nonlinearity parameter $\mu$ in our models can then be interpreted as the strength of the feedback in-loop measurement. It should however be noted that the measurement needs to be invasive, or direct. Therefore the measurement in Appendices~\ref{InterpretRayleigh} and \ref{vdPApp} cannot be interpreted using conventional quantum-optical techniques because such conventional methods only measure the output of the system, not the system directly.

\subsection{Limit-cycle behaviour}

\subsubsection{Rayleigh oscillator}
\label{RayleighW(Q,P)}

\begin{figure}[t]
\centerline{\includegraphics[width=0.42\textwidth]{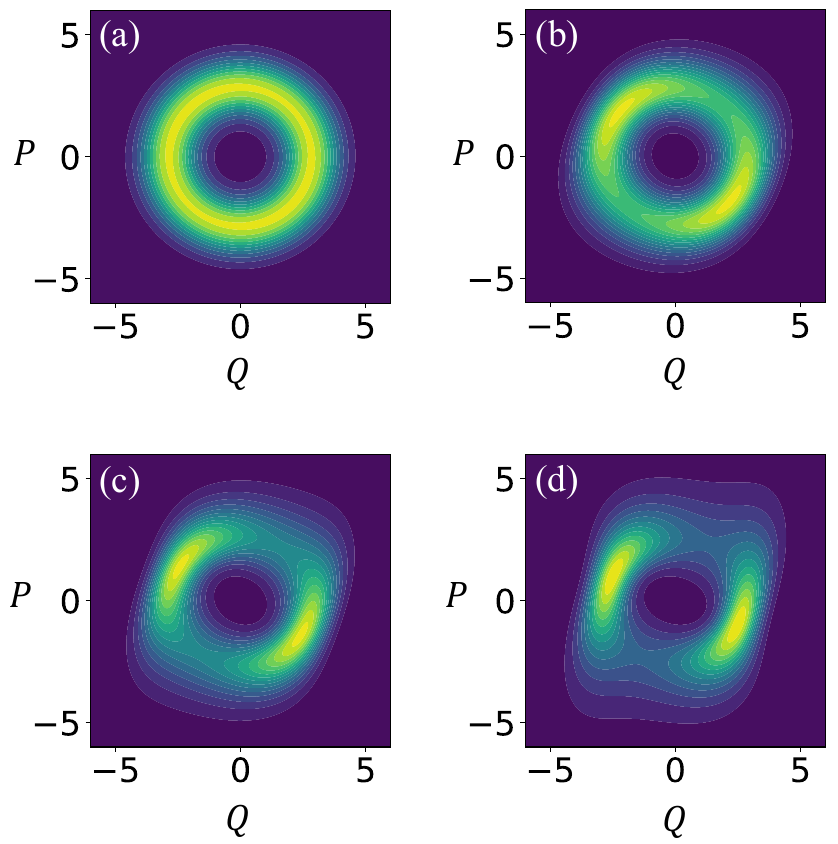}}
\caption{\label{QvdPLimitCycles} Steady-state Wigner functions for the Rayleigh oscillator with $q_0=3$ and $\wo=1$. The axes are $Q=q/2$ (horizontal) and $P=p/2$ (vertical). The contour heights are represented by yellow, green and blue in descending order: (a) $\mu=0.001$. (b) $\mu=0.02$. (c) $\mu=0.05$. (d) $\mu=0.1$.}
\end{figure}

The notion of a limit cycle for a quantum system is captured by its steady-state Wigner function so we will also refer to it loosely as a quantum limit cycle (keeping in mind that the steady-state Wigner function is static, unlike the trajectory of a classical phase-space point). Let us denote the steady-state Wigner function for $\Lcal_\sqr$ by $W_\sqr(Q,P)$, which we note is a positive function everywhere in phase space and is therefore a true probability density. Note here that we are using the scaled variables $Q=q/2$ and $P=p/2$. This is shown in Fig.~\ref{QvdPLimitCycles}\,(a)--(d) for increasing strengths of nonlinearity (see figure caption for parameter values). Figure~\ref{QvdPLimitCycles}\,(a) shows that $\Lcal_\sqr$ recovers a circular $W_\sqr(Q,P)$ which one expects for $\mu\longrightarrow 0$ \cite{LS13,WNB14}. Note the circularity seen in Fig.~\ref{QvdPLimitCycles}\,(a) is not a trivial result. This is because every term in $\Lcal_\sqr$ has $\mu$, including the ones which also appear in $\Lcal_\smcirc$. The convergence to a circular limit cycle for $\mu\too0$ is therefore reassuring.

As we increase $\mu$ from Fig.~\ref{QvdPLimitCycles}\,(a), the shape of $W_\sqr(Q,P)$ starts to deviate from a circle and becomes more square-like just as the classical limit cycle would. However, in contrast to our analysis based on the $q$ nullcline in Sec.~\ref{RayClass}, Fig.~\ref{QvdPLimitCycles} does not show an extension of the quantum limit cycle along the $P$ direction as $\mu$ gets larger. Instead, we find that as $\mu$ increases, $W_\sqr(Q,P)$ becomes more bimodal. Hence we see the first interesting departure from the $\mu \too 0$ limit cycle: The bimodality of $W_\sqr(Q,P)$ shows that, at large $\mu$, it cannot simply be thought of as the classical limit cycle with some width added to it to account for quantum noise. Nevertheless, it is qualitatively consistent with relaxation oscillations. This is because in a region of phase space where we expect the oscillator to be moving slowly, we also expect a high probability of finding the oscillator there. Conversely, we expect a low probability of finding the oscillator in a phase-space region where it is moving quickly. To see this, we can imagine an ensemble of classical relaxation oscillators, possibly all with different initial conditions but allowed to settle to the limit cycle. At large $\mu$, each oscillator only spends a tiny fraction of its period in the fast sections of the limit cycle while the remaining time is spent along the slow sections. If we were to pick a random member of this ensemble then we will most likely find its phase-space point in one of the slow sections. It therefore makes sense that the high and low regions of $W_\sqr(Q,P)$ correspond to where the classical limit cycle is expected to be slow and fast respectively [see Fig.~\ref{RayLC}\,(a)]. The association of $W_\sqr(Q,P)$ to the time spent along different sections of the corresponding classical limit cycle is also qualitatively consistent with an analogous classical nonlinear oscillator with additive noise seen in Refs.~\cite{TS85,TS86,KSSG03}. With this understanding we can now see why $\mu\too 0$ produces a steady-state Wigner function with equal probability distributed around a circle---because in this limit, the corresponding classical limit cycle is traversed with a uniform speed and therefore spends the same amount of time everywhere on its limit cycle. 

Other than the elongation of the limit cycle, our analysis of the $q$ nullcline in Sec.~\ref{RayClass} also predicts a slight tilt in the limit cycle as $\mu$ increases (for $q_0$ and $\wo$ constants). This effect can now be seen in Fig.~\ref{QvdPLimitCycles}. Our interpretation of the Wigner-function peaks now permits us to think of the quantum limit cycle in Fig.~\ref{QvdPLimitCycles}\,(b) as slanting to the right. Then as we increase $\mu$, going from Fig.~\ref{QvdPLimitCycles}\,(b) to (c), the quantum limit cycle tilts slightly towards the left. This happens again in going from Fig.~\ref{QvdPLimitCycles}\,(c) to (d), qualitatively consistent with what we expect from Sec.~\ref{RayClass}.

\begin{figure}[t]
\centerline{\includegraphics[width=0.38\textwidth]{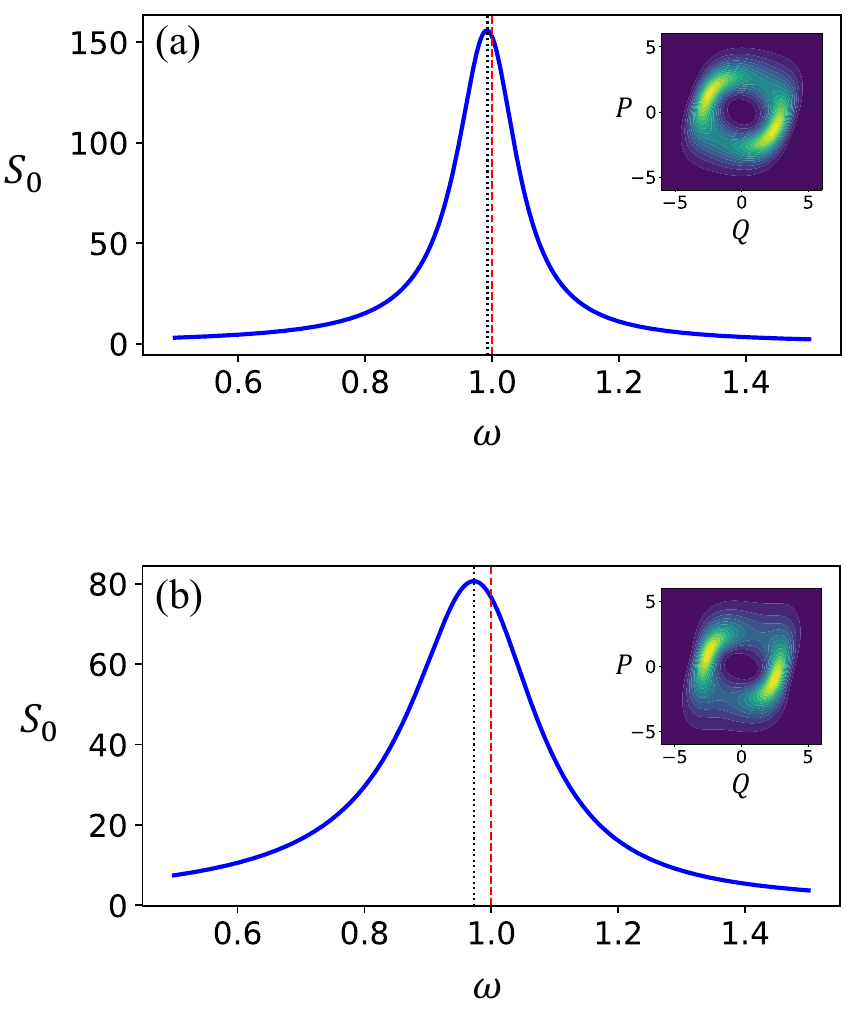}}
\caption{\label{RayleighFreqShift} Spectra of the Rayleigh oscillator for $q_0=2$ and $\wo=1$ and the corresponding quantum limit cycles (insets). The Rayleigh oscillator's frequency at $\mu=0$ is given by $\wo$ and is indicated by the vertical red dashed line. Its frequency for nonzero $\mu$, denoted by $\Omega_0$, is indicated by the vertical black dotted line. (a) $\mu=0.05$, $\Omega_0=0.992$. (b) $\mu=0.1$, $\Omega_0=0.972$.}
\end{figure}

Another departure from the weakly-nonlinear limit which has not been observed is the shift in the oscillator's frequency away from $\wo$ when $\mu$ is allowed to vary. Though the relationship of this shift to the oscillator's nonlinearity can be expected to be difficult to deduce in quantum mechanics, our model at least facilitates a numerical investigation. To this end let us denote the oscillator's frequency for nonzero $\mu$ by $\Omega_0$. This is defined by
\begin{align}
\label{Omega0ss}
	\Omega_0 = \text{argmax} \;  S_0(\omega)  \;  ,
\end{align}
where $S_0(\omega)$ is the oscillator's steady-state power spectrum. The subscript zero is meant to remind us that the oscillator is free of any external forcing, so $\Omega_0$ is its natural frequency. It is given by the Fourier transform of the correlation function for the operator $\delta\ann(t) = \ann(t) - \an{\ann(t)}$:
\begin{align}
\label{S(w)ss}
	S_0(\omega) = \int^{\infty}_{-\infty} d\tau \; e^{i\omega\tau} \, \ban{\delta\adg(0)\, \delta\ann(\tau)}  \; ,
\end{align}
where all expectation values are taken respect to the steady state of $\Lcal_\sqr$ [and hence $\an{\ann(t)}$ is in fact independent of $t$]. We show  frequency shifts experienced by the Rayleigh oscillator for $\mu=0.05$ in Fig.~\ref{RayleighFreqShift}\,(a), and $\mu=0.1$ in Fig.~\ref{RayleighFreqShift}\,(b), both corresponding to $q_0=2$ and $\wo=1$. Their spectra are shown as blue solid lines and the corresponding quantum limit cycles are shown as insets. We find the oscillator's frequency does indeed change when we increase $\mu$. Note the frequency shift is towards the left of the red dashed line, i.e.~a decrease from $\wo$, and the shift is greater for a greater $\mu$. This is consistent with the classical theory of relaxation oscillations in strongly-nonlinear oscillators \cite{Str15,Sco05}. Interestingly, the insets of Fig.~\ref{RayleighFreqShift} show that even though the quantum limit cycles appear to have deviated from the weakly-nonlinear regime, the oscillator's frequency as defined by \eqref{Omega0ss}, has only shifted very slightly. Thus the amount of frequency shift as a function of $\mu$ requires a more detailed study in quantum mechanics.

\subsubsection{\vdp\ oscillator}

\begin{figure}[t]
\centerline{\includegraphics[width=0.42\textwidth]{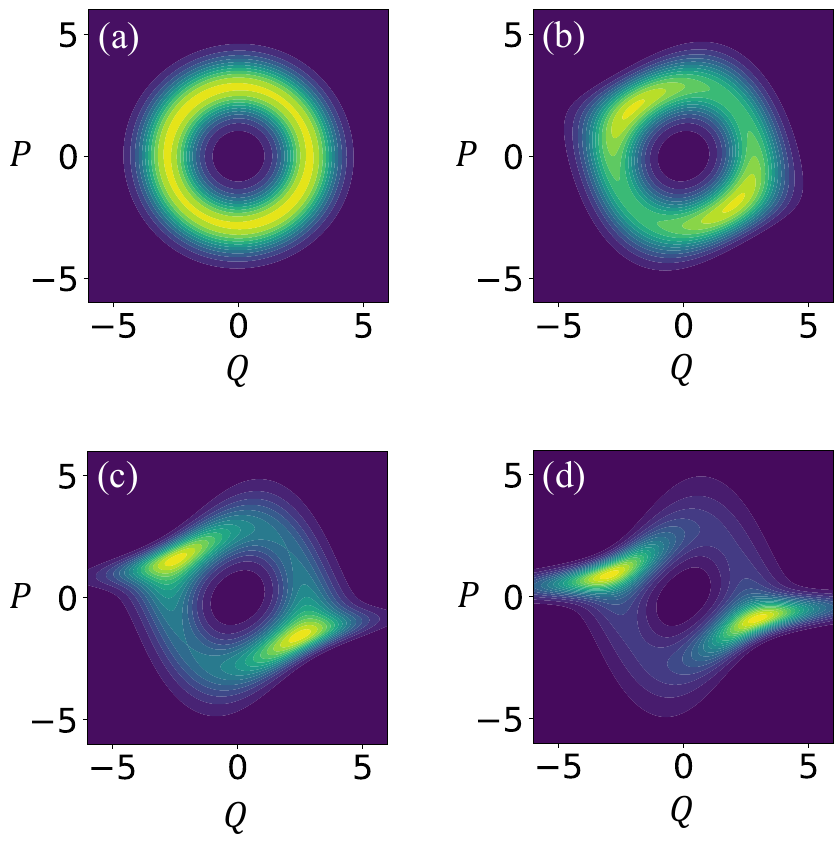}}
\caption{\label{QvdPLimitCyclesDiamond} Steady-state Wigner function of the \vdp\ oscillator for $q_0=3$ and $\wo=1$. The axes are $Q=q/2$ (horizontal) and $P=p/2$ (vertical). The contour heights are represented by yellow, green and blue in descending order. (a) $\mu=0.001$. (b) $\mu=0.02$. (c) $\mu=0.05$. (d) $\mu=0.1$.}
\end{figure}

If the above interpretation of the Rayleigh oscillator's steady-state Wigner function is correct, then it should also apply to the \vdp\ oscillator. Recall from Sec.~\ref{ClassvdP} that we showed the slow sections of the \vdp\ limit cycle to be in quadrants two and four of phase space, one of which is boxed in Fig.~\ref{vdPFlow}. In Fig.~\ref{QvdPLimitCyclesDiamond} we show the steady-state \vdp\ Wigner function $W_\dia(Q,P)$ for varying strengths of nonlinearity. As we increase $\mu$ from Fig.~\ref{QvdPLimitCyclesDiamond}\,(a) to (d) we find the location of the peaks of $W_\dia(Q,P)$ corroborate with the slow portions of the classical limit cycle seen in Sec.~\ref{ClassvdP}. But unlike the classical case, these parts of the quantum limit cycle extend outwards quite significantly along the direction of $Q$ which can be seen in Fig.~\ref{QvdPLimitCyclesDiamond}\,(d).

We can see from Fig.~\ref{QvdPLimitCyclesDiamond}\,(a) that the \vdp\ oscillator also has the expected weakly-nonlinear limit. Having two distinctly different limit cycles both become circular provides evidence that $\Lcal_\smcirc$  may be regarded as a universal model  in some sense for $\mu\too0$ [see the discussion surrounding \eqref{QSL} and \eqref{2ndOrder} in Sec.~\ref{Intro}]. It might be thought that $\Lcal_\mu\too\Lcal_\smcirc$ as $\mu\too0$, where $\Lcal_\mu$ may be $\Lcal_\sqr$ or $\Lcal_\dia$. However, this is too demanding because it is unreasonable to expect the quantum noise in $\Lcal_\mu$ to behave exactly like the quantum noise in $\Lcal_\smcirc$, even for small $\mu$. This in turn is because the quantum models represented by $\Lcal_\mu$ are only constrained by the mean of $\ann$ and nothing else. As we mentioned in Sec.~\ref{GeneralRemarks}, it is precisely this lack of constraint in the definition of quantum dynamical systems that leads to their non-uniqueness. Likewise, the quantum Stuart--Landau model defined by \eqref{QSL} in Sec.~\ref{Intro} is also not unique. Reference~\cite{CHN+19} provides an explicit example where different quantum noises corresponding to different quantisations of the same system lead to an interesting consequence. Thus, in terms of the moments of $\ann$, all we can demand from $\Lcal_\smcirc$ is that
\begin{align}
\label{LimitInTheMean}
	\Lcal_\mu  \; \stackrel{\an{\ann}}{\too} \;  \Lcal_\smcirc   \; ,   \quad \mu \too 0  \; .
\end{align}
The symbol $\stackrel{\an{\ann}}{\too}$ in \eqref{LimitInTheMean} denotes that $\Lcal_\mu$ behaves like $\Lcal_\smcirc$ only in $\an{\ann}'$.

We may also think about the universality of $\Lcal_\smcirc$ in terms of the quantum limit cycles that we have been plotting. If $\Lcal_\smcirc$ is a universal model for $\mu\too0$, then we can expect there to be some $\gamma_1$ and $\gamma_2$ in \eqref{QSL} such that the mode of its steady-state Wigner function $W_\smcirc(Q,P)$, matches the mode of $W_\sqr(Q,P)$ for a given $\wo$, $q_0$ in $\Lcal_\sqr$. Given our discussion about quantum noise in the preceding paragraph, other features of  the two quantum limit cycles given by $W_\smcirc(Q,P)$ and $W_\sqr(Q,P)$ cannot be expected to match (except for the average $Q$ and average $P$, which are zero). The same applies to $W_\smcirc(Q,P)$ and $W_\dia(Q,P)$ for a given $\wo$ and $q_0$ in the quantum \vdp\ oscillator.

\section{Driven quantum oscillator}
\label{Qsync}

The question now is if the shark-fin like waveforms characteristic of relaxation oscillations seen in Fig.~\ref{RayLC}\,(b) and Fig.~\ref{vdPLC}\,(b) of Sec.~\ref{Class_Ray+vdP} are contained in our quantum models of the Rayleigh and \vdp\ oscillators. The technical obstacle to overcome in this problem is how one can sensibly introduce the time dependence needed for these waveforms to be extracted \footnote{One potential method is to continuously measure the oscillator. We found that a continuous measurement of the oscillator's amplitude via heterodyning produces only a very faint evidence of relaxation behaviour. Furthermore, this does not produce smooth curves for the oscillations because the measurement results are noisy.}. Since we are also interested in frequency entrainment (Sec.~\ref{Entrainment}) we can simply use the drive as a means of extracting relaxation oscillations (Sec.~\ref{PhaseSpaceSlime}).

\subsection{Frequency entrainment}
\label{Entrainment}

For simplicity we will consider frequency entrainment only for $\Lcal_\sqr$\,. The driven oscillator model is thus
\begin{align}
\label{vdPDriven}
	\rho' = -i \, \epsilon \,\cos(\omega_1 \, t + \phi) \, \big[ \ann+\adg, \rho \big] + \Lcal_\sqr \, \rho  \;.
\end{align}
Here $\epsilon$ is the drive's strength, $\omega_1$ its frequency, and $\phi$ an arbitrary initial phase. As we alluded to in Sec.~\ref{GeneralRemarks}, there is no frame in which the drive's time dependence can be removed, and this is independent of the form of the oscillator-drive coupling. For a classical oscillator, entrainment in this case can be demonstrated stroboscopically. This means observing the oscillator's $q$ and $p$ at the instants $t_n \equiv t_0 + n \, T$ where $T = 2\pi/\omega_1$ and $n$ can be any positive integer with $t_0 \rightarrow \infty$. If the oscillator is entrained to the drive, then wherever the phase-space point appears at $t_0$, it should also be found at the exact same point in phase space at integer multiples of the drive's period thereafter.

For a quantum oscillator, quantum noise ruins the possibility of a precisely defined phase-space point. So all one can expect for a quantum oscillator is that snapshots of its Wigner function be localised in the same phase-space region at the stroboscopic times $t_n$. In Appendix~\ref{FidApp} we verify that \eqref{vdPDriven} produces a periodic $\rho(t)$ at the drive's frequency. This implies that the time-dependent Wigner overlaps with itself at $t_n$. However, the periodicity of $\rho(t)$ does not imply frequency entrainment because the quantum noise formally encapsulated in $\rho(t)$ prevents the oscillator from attaining a frequency equal to $\omega_1$. It has already been shown in the weakly-nonlinear regime that quantum noise in fact makes frequency entrainment quite imperfect \cite{WNB14}.

If the oscillator entrains to the external drive, its frequency should shift towards $\omega_1$ as $\epsilon$ is increased for a fixed nonlinearity and a fixed detuning $\Delta=|\omega_1-\Omega_0|$. Recall from \eqref{Omega0ss} and \eqref{S(w)ss} that we have defined the oscillator's frequency by where its spectrum peaks. What we need here is the frequency of the driven oscillator. But as we explained, the external drive's time dependence cannot simply be dealt with by moving into a rotating frame for a relaxation oscillator. Hence, the correlation function $\an{\delta\adg(t) \, \delta\ann(t+\tau)}$, where $\delta\ann(t)=\ann(t)-\an{\ann(t)}$, may now depend on the time origin $t$, in addition to $\tau$. This leads to a time-dependent spectrum
\begin{align}
\label{S(w,t)}
	S_1(\omega,t) = \int^\infty_{-\infty} d\tau \; e^{i\omega \tau} \, \ban{\delta\adg(t) \, \delta\ann(t+\tau)}  \; .
\end{align}
To extract the driven oscillator's frequency we will consider its spectrum averaged over one period of the drive \cite{NN19}. That is, we define the frequency of a driven oscillator to be
\begin{align}
	\Omega_1 = {\rm argmax} \; \overline{S}_1(\omega)   \; ,
\end{align}
where now
\begin{align}
\label{Sbar(w)}
	\overline{S}_1(\omega) = \frac{1}{T} \int^{t_0+T}_{t_0} \, dt'  \;  S_1(\omega,t')  \; .  
\end{align}  
In practice a discrete but sufficiently large set of $t$ values are used to calculate the time-averaged spectrum, and \eqref{Sbar(w)} may be approximated by
\begin{align}
	\overline{S}_1(\omega) = \frac{1}{M} \; \sum_{k=1}^M  \; S_1(\omega,t_k)  \; ,  \quad  t_k \in [\,t_0,t_0+T\,]   \; .
\end{align}  
\begin{figure}[t]
\centerline{\includegraphics[width=0.38\textwidth]{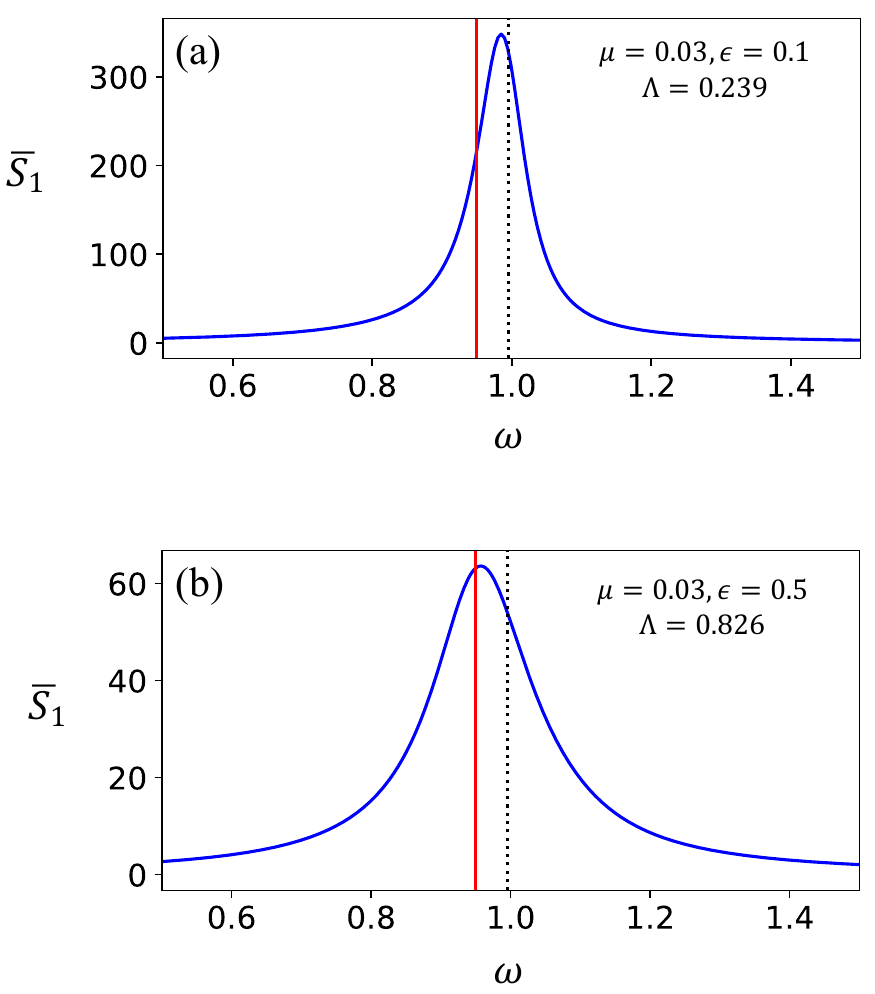}}
\caption{\label{SyncSmallmu} Frequency shifts of the Rayleigh oscillator at different driving strengths for $\mu=0.03$, $q_0=2$, $\wo=1$, $\omega_1=0.95$  (vertical red solid line), $\phi=\pi/2$. Numerical results are rounded up to three decimal places. These parameter values lead to $\Omega_0=0.996$ (vertical black dotted line) and $\Delta=0.046$. (a) $\epsilon=0.1$, $\Lambda=0.239$. (b) $\epsilon=0.5$, $\Lambda=0.826$.}
\end{figure}
In practice we find the driven Rayleigh oscillator has a $\an{\delta\adg(t) \, \delta\ann(t+\tau)}$ that depends on $t$ only very weakly so that we can approximate $\delta\ann(t)$ as a second-order stationary quantum stochastic process and use $S_1(\omega,t)$ instead of $\overline{S}_1(\omega)$ if one wishes. This is illustrated in Appendix~\ref{CorrFuncApp}. We also require a sufficiently large value of $t_0$ in \eqref{Sbar(w)} to avoid transient effects in the power spectrum. This is to allow the external drive sufficient time to entrain the oscillator (see Appendix~\ref{FidApp} for further details). We then take, as a measure of frequency entrainment, the amount of frequency shift under driving normalised by the detuning,
\begin{align}
	\Lambda = \frac{|\Omega_1 - \Omega_0|}{\Delta}   \; .
\end{align} 
If there is no frequency entrainment we expect $\Omega_1=\Omega_0$ and $\Lambda=0$. For perfect entrainment, $\Omega_1=\omega_1$ and $\Lambda=1$.

\begin{figure}[t]
\centerline{\includegraphics[width=0.38\textwidth]{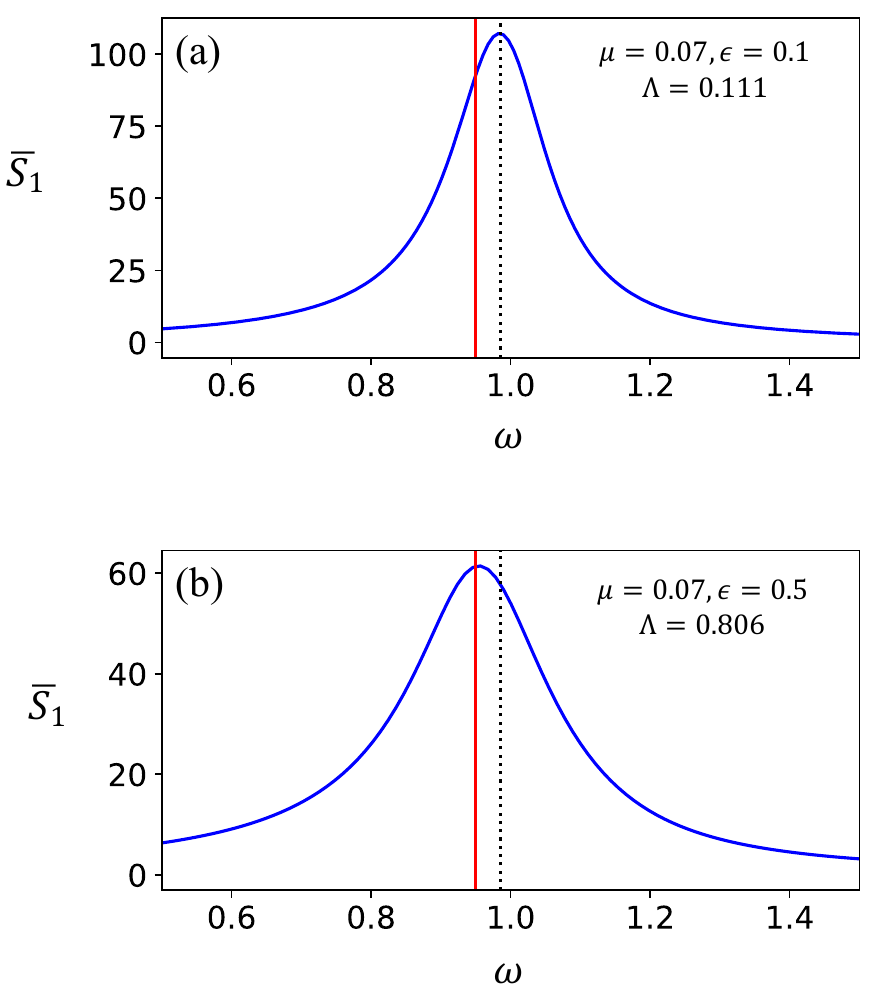}}
\caption{\label{SyncLargemu} Frequency shifts of the Rayleigh oscillator at different driving strengths for $\mu=0.07$, $q_0=2$, $\wo=1$, $\omega_1=0.95$ (vertical red solid line), $\phi=\pi/2$. Numerical results are rounded up to three decimal places. These parameter values lead to $\Omega_0=0.986$ (vertical black dotted line) and $\Delta=0.036$. (a) $\epsilon=0.1$, $\Lambda=0.111$. (b) $\epsilon=0.5$, $\Lambda=0.806$.}
\end{figure}

We now illustrate frequency entrainment for two different nonlinearities. In Fig.~\ref{SyncSmallmu} for $\mu=0.03$, and in Fig.~\ref{SyncLargemu} for $\mu=0.07$ (see figure captions for other parameter values). In both figures, the spectrum of the driven oscillator corresponds to the blue solid line while the frequencies $\omega_1$ (of the drive), and $\Omega_0$ (of the undriven oscillator), are shown as the red solid and black dotted lines respectively.

In both cases we consider $\Lambda$ for different driving strengths at $\epsilon=0.1$ [Figs.~\ref{SyncSmallmu} and \ref{SyncLargemu}\,(a)], then $\epsilon=0.3$ (spectra not shown), and finally $\epsilon=0.5$ [Figs.~\ref{SyncSmallmu} and \ref{SyncLargemu}\,(b)]. For $\epsilon=0.1$ we find entrainment to be rather weak as shown by the respective $\Lambda$ values in the inset and figure captions \footnote{Our frequency axis has a spacing of $\Delta\omega=0.007$ which sets an upper bound on the accuracy of our estimate for $\Omega_1$ and hence $\Lambda$.  In particular, the numerical estimate of $\Lambda$ in Fig.~\ref{SyncLargemu}\,(a) is based on $\Omega_1 = 0.982$, which lies within $[\Omega_0-\Delta\omega,\Omega_0+\Delta\omega]$. This exceptional case aside, the other data points, and the general conclusion on frequency entrainment drawn from them, are valid.}. Increasing the driving strength to $\epsilon=0.3$ we find that for $\mu=0.03$, $\Lambda=0.348$, and for $\mu=0.07$, $\Lambda=0.500$, indicating a slight improvement in frequency entrainment. 
An appreciable difference occurs when we further increase the driving strength to $\epsilon=0.5$. This can be seen in the $\Lambda$ values corresponding to the two different nonlinearities in Figs.~\ref{SyncSmallmu} and \ref{SyncLargemu}. That a stronger drive improves frequency entrainment has been known for the quantum Stuart--Landau oscillator and here we are simply seeing an extension of this effect to a strongly-nonlinear quantum oscillator.

\subsection{Relaxation oscillations}
\label{PhaseSpaceSlime}

If we would like to observe relaxation oscillations in the quantum Rayleigh oscillator then we must find suitable statistics that can give us a meaningful correspondence to the oscillator's position and momentum. We find the mode provides this correspondence, as opposed to the mean, which might have been one's first guess. That is, for sufficiently large $\mu$, relaxation oscillations can be found in
\begin{align}
\label{ModeDefn}
	\big( Q_\star(t), P_\star(t) \big) = {\rm argmax} \; \tilde{W}_\sqr(Q,P,t)  \; , 
\end{align}
where $\tilde{W}_\sqr(Q,P,t)$ is the Wigner function for the driven oscillator. In \eqref{ModeDefn} we treat $t$ as a parameter. 

We shall see below that relaxation oscillations  can be broadly classified into two types, depending on the relative strength of the oscillator's nonlinearity $\mu$ compared to the driving strength $\epsilon$.

\subsubsection{Unimodal behaviour}

\begin{figure}[t]
\centerline{\includegraphics[width=0.35\textwidth]{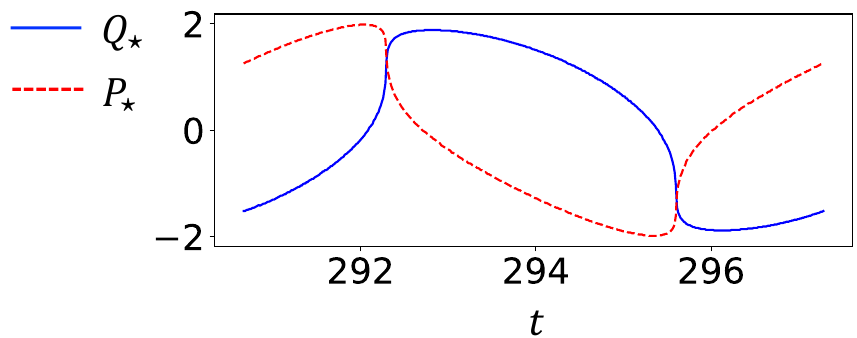}}
\caption{\label{RelaxOscUnimodal} Relaxation oscillations in $Q_\star(t)$ (blue solid line) and $P_\star(t)$ (red dashed line) for one period of the drive starting at $t_0=8.72/\mu=290.66$. This time is chosen to allow the driven oscillator to reach its long-time limit (see Appendix~\ref{FidApp}) and to make the shark fin appear in the centre of the time window. The parameter values are identical to Fig.~\ref{SyncSmallmu}\,(a):  $\mu=0.03$, $q_0=2$, $\wo=1$, $\Omega_0=0.996$, $\omega_1=0.95$, $\phi=\pi/2$, and $\epsilon=0.1$.}
\end{figure}

We focus first on the case when the oscillator's nonlinearity can be considered weak compared to the drive. A suitable choice of parameter values is $\mu=0.03$ and $\epsilon=0.1$, with all other parameters the same as in Fig.~\ref{SyncSmallmu}. In this case the oscillator's frequency is weakly entrained to the drive. Its associated $(Q_\star,P_\star)$ is plotted in Fig.~\ref{RelaxOscUnimodal}. They reveal relaxation oscillations in the quantum Rayleigh oscillator\;\!! Unlike the classical Rayleigh oscillator, its quantum analogue shows relaxation oscillations in both the position and momentum. The waveform in Fig.~\ref{RelaxOscUnimodal} persists indefinitely so long as the external drive is there to sustain it although we have only displayed one period of it here. We have already noted above that \eqref{vdPDriven} has a periodic solution whose period is $T=2\pi/\omega_1$ (see Appendix~\ref{FidApp}). This implies that $\tilde{W}_\sqr(Q,P,t)$ will be periodic. Hence all its statistics are periodic, including its mode.

\begin{figure}[t]
\centerline{\includegraphics[width=0.48\textwidth]{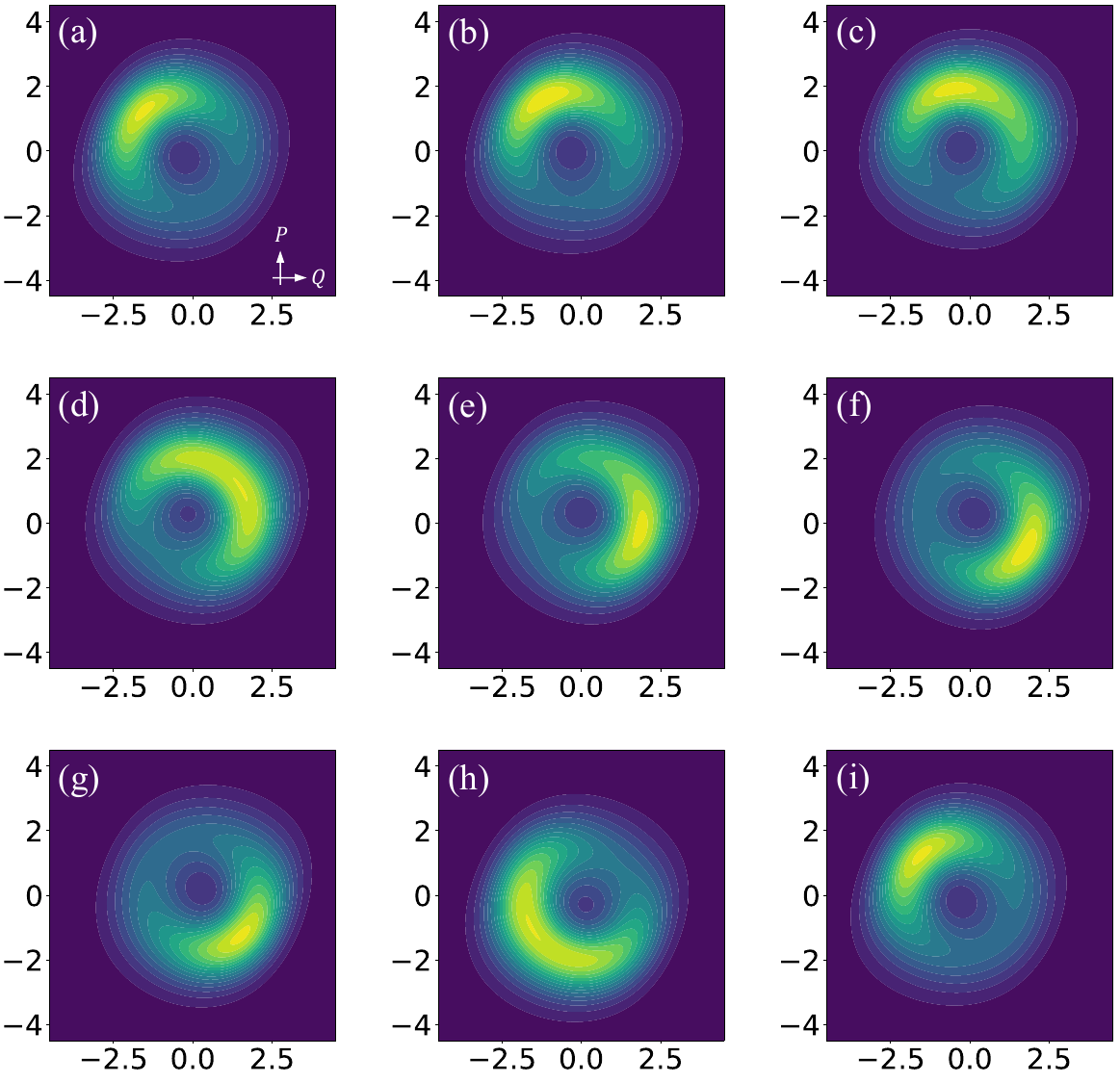}}
\caption{\label{QuantumSlimeUni} Wigner functions corresponding to Fig.~\ref{RelaxOscUnimodal}. Horizontal axis is $Q$ and the vertical axis is $P$ [see inset of (a)]. (a)--(g) Seven snapshots $\tilde{W}_\sqr(Q,P,t)$ for $t \in [t_0,t_0+T/2]$, divided evenly into six intervals. (h) $\tilde{W}_\sqr(Q,P,t_0+3T/4)$. (i) $\tilde{W}_\sqr(Q,P,t_0+T)$.}
\end{figure}

Looking at Fig.~\ref{RelaxOscUnimodal} we see that the mode is a representation of ``where the particle is'' at each moment in time. Of course the complete dynamics of the quantum oscillator cannot simply be reduced to $(Q_\star,P_\star)$, since there is also quantum noise. Thus, a more complete understanding of how relaxation oscillations should refer to the entire Wigner function $\tilde{W}_\sqr(Q,P,t)$, not just its peak. We now examine Fig.~\ref{RelaxOscUnimodal} in more detail using Fig.~\ref{QuantumSlimeUni} where we plot the Wigner function corresponding to nine different times along the relaxation oscillation as follows. Assuming the waveform in Fig.~\ref{RelaxOscUnimodal} starts at $t_0$, and ends at $t_0+T$, we divide $[t_0,t_0+T]$ into two segments, from $t_0$ to $t_0+T/2$, and then from $t_0+T/2$ to $t_0+T$: We further divide $[t_0,t_0+T/2]$ into six equally spaced intervals. This is shown in Fig.~\ref{QuantumSlimeUni}\,(a)--(g). The remaining interval $[t_0+T/2,t_0+T]$ is then halved, corresponding Fig.~\ref{QuantumSlimeUni}\,(g)--(i).

In Fig.~\ref{QuantumSlimeUni}\,(a) we catch the oscillator when it is in transit through its slow section on the limit cycle. We can also see from $(Q_\star,P_\star)$ in Fig.~\ref{RelaxOscUnimodal} that the oscillator is approaching the fast section of the limit cycle. Figure~\ref{QuantumSlimeUni}\,(a)--(g) gives us a breakdown of just how this occurs in phase space. We find in Fig.~\ref{QuantumSlimeUni}\,(a)--(c) that the oscillator is holding onto its position along the slow section of the limit cycle. The Wigner function then starts to diffuse to the fast section. Hence the tendency to transit the fast section of the limit cycle starts to grow which can be seen in Fig.~\ref{QuantumSlimeUni}\,(b)--(d). In Fig.~\ref{QuantumSlimeUni}\,(d)--(e) the oscillator finally gives in and transits the fast side of the limit cycle. It then reaches the other slow side in Fig.~\ref{QuantumSlimeUni}\,(f) and crawls along it in Fig.~\ref{QuantumSlimeUni}\,(f)--(g). This process of diffuse-and-zap then repeats itself in Fig.~\ref{QuantumSlimeUni}\,(g)--(i). This behaviour in phase space is reminiscent of the slimy behaviour seen in the sticky-hand toy played by children \footnote{see Google images for ``sticky hand toy''.}.

We can also understand the phase-space behaviour qualitatively by recalling from Fig.~\ref{QvdPLimitCycles} that a strongly-nonlinear oscillator has a bimodal steady-state Wigner function (which is the parameter regime of Fig.~\ref{QuantumSlimeUni} for $\epsilon=0$). The external drive can be represented by a phasor in phase space whose effect is to force the oscillator's phase-space distribution to localise around its tip. However, we can expect this to be possible only when the drive is sufficiently strong to overpower the oscillator's intrinsic bimodality. That is, when the oscillator's nonlinearity is weak compared to the driving strength, the oscillator produces a unimodal $\tilde{W}_\sqr(Q,P,t)$ in phase space and this is what we are seeing in Fig.~\ref{QuantumSlimeUni}.

\subsubsection{Bimodal behaviour}

\begin{figure}[t]
\centerline{\includegraphics[width=0.35\textwidth]{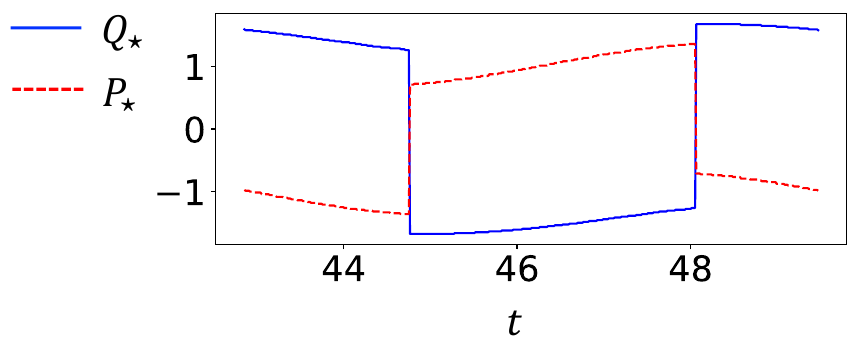}}
\caption{\label{RelaxOscBimodal} Relaxation oscillations in $Q_\star(t)$ (blue solid line) and $P_\star(t)$ (red dashed line) for one period of the drive starting at $t_0=3/\mu=42.85$ (chosen for the same reasons as in Fig.~\ref{RelaxOscUnimodal}). The parameter values are almost the same as Fig.~\ref{SyncLargemu}\,(a), just with a weaker drive: $\mu=0.07$, $q_0=2$, $\wo=1$, $\Omega_0=0.986$, $\omega_1=0.95$, $\phi=\pi/2$, and $\epsilon=0.06$.}
\end{figure}

To see what happens when the oscillator becomes more nonlinear and go into a regime where its bimodality dominates the drive, we keep the same set of parameter values as in Figs.~\ref{RelaxOscUnimodal} and \ref{QuantumSlimeUni}, but now let $\mu=0.07$ and $\epsilon=0.06$. In this case there is no observable frequency entrainment. The resulting $Q_\star(t)$ and $P_\star(t)$ is shown in Fig.~\ref{RelaxOscBimodal} where as before, it is shown over one period of the drive starting at $t_0$. In this case we see that $Q_\star(t)$ and $P_\star(t)$ are nearly square waves. That is, the slow and fast timescales of the relaxation oscillations have become extremely different. 

\begin{figure}[t]
\centerline{\includegraphics[width=0.48\textwidth]{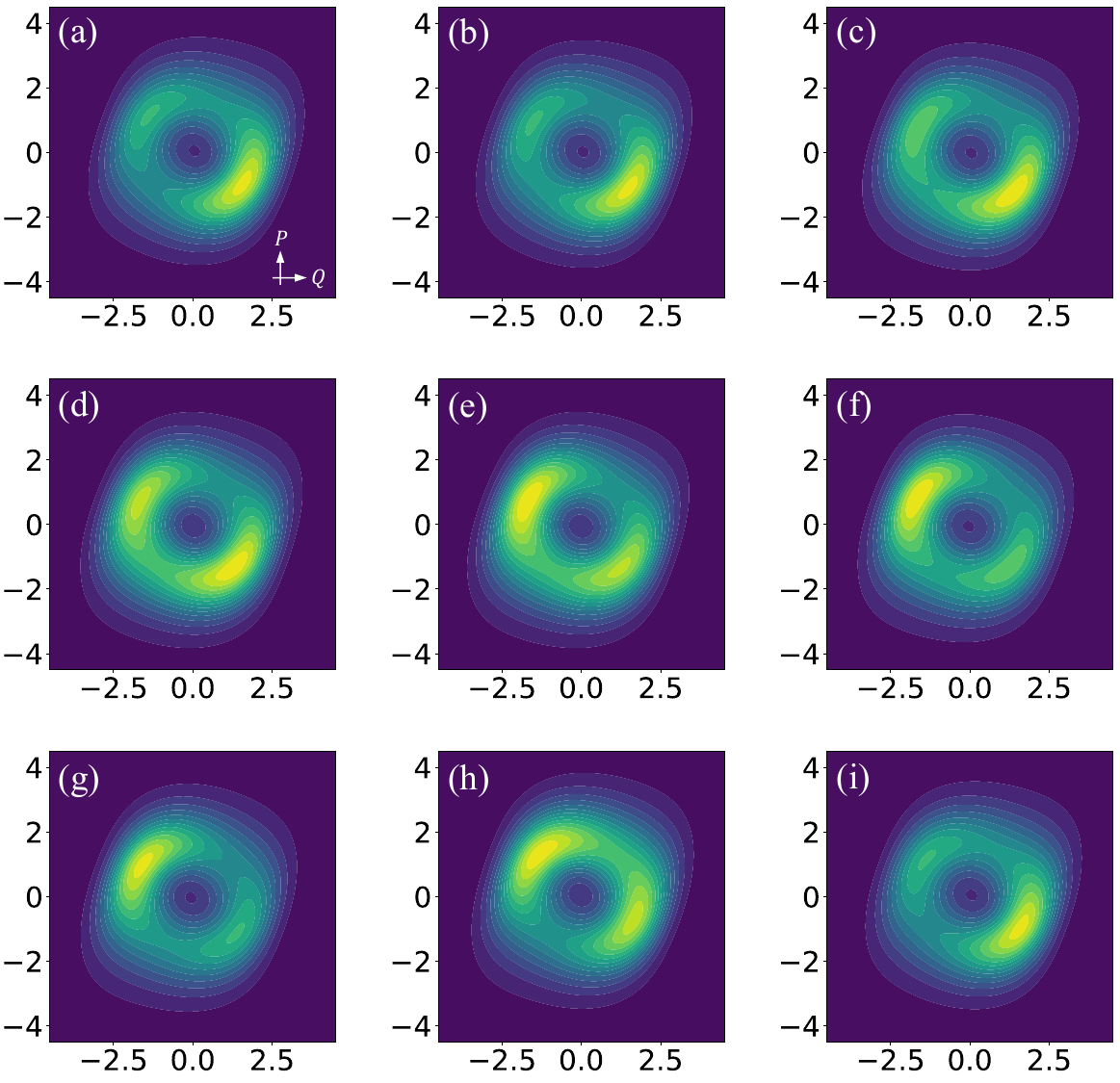}}
\caption{\label{QuantumSlimeBi} Wigner functions corresponding to Fig.~\ref{RelaxOscBimodal}.  Horizontal axis is $Q$ and the vertical axis is $P$ [see inset of (a)]. (a)--(g) Seven snapshots $\tilde{W}_\sqr(Q,P,t)$ for $t \in [t_0,t_0+T/2]$, divided evenly into six intervals.  (h) $\tilde{W}_\sqr(Q,P,t_0+3T/4)$. (i) $\tilde{W}_\sqr(Q,P,t_0+T)$.  A slight amount of diffusion is noticeable in one of the fast sections of the quantum limit cycle when the peak of $\tilde{W}_\sqr(Q,P,t)$ is being transferred [in (d), and again in (h)]. Hence there is still a preferred direction of travel but the amount of diffusion here is incomparable to the unimodal case in Fig.~\ref{QuantumSlimeUni}.}
\end{figure}

To explain Fig.~\ref{RelaxOscBimodal} we again consider $\tilde{W}_\sqr(Q,P,t)$ at intermediate times between $t_0$ and $t_0+T$ and break up the interval $[t_0,t_0+T]$ as we did in Fig.~\ref{QuantumSlimeUni}. The snapshots of the Wigner functions are shown in Fig.~\ref{QuantumSlimeBi}. As in the unimodal case, we start with the oscillator at its slow section of the limit cycle at $t_0$. As time progresses the relaxation oscillations maintain a nearly constant amplitude for a while. This corresponds to Fig.~\ref{QuantumSlimeBi}\,(a)--(c). As we approach the first discontinuity in Fig.~\ref{RelaxOscBimodal}, the Wigner function does not diffuse like the unimodal case [compare Fig.~\ref{QuantumSlimeUni}\,(c)--(d) to Fig.~\ref{QuantumSlimeBi}\,(c)--(d)]. Rather, the Wigner function at the site of the mode begins to diminish while it builds up on the opposite slow side of the limit cycle. We catch this in action in Fig.~\ref{QuantumSlimeBi}\,(d) and (e). Here we see the peak of $\tilde{W}_\sqr(Q,P,t)$ being transferred between the two slow sides of the limit cycle without it having to trace through the fast section. This explains the discontinuity seen in Fig.~\ref{RelaxOscBimodal}: At some point in time the peak of the Wigner function on one side of the limit cycle becomes just marginally higher than the opposing side, and that is when the discontinuity occurs in Fig.~\ref{RelaxOscBimodal}. Once it is on the other slow side as seen in Fig.~\ref{QuantumSlimeBi}\,(f) and (g), the Wigner function then only moves a little bit which accounts for the slope of $Q_\star(t)$ and $P_\star(t)$ in Fig.~\ref{RelaxOscBimodal}. This entire process then repeats again in going from Fig.~\ref{QuantumSlimeBi}\,(g) to (i).

Thus we see that in the unimodal regime the oscillator effects relaxation oscillations by executing a sequence of diffuse-and-zap transitions in phase space, while in the bimodal regime the oscillator disappears from one side of the limit cycle and reappears on the other. We can also contrast the unimodal regime with the bimodal one on a longer timescale by focusing only on Fig.~\ref{QuantumSlimeUni}\,(g)--(i)  and Fig.~\ref{QuantumSlimeBi}\,(g)--(i).

\section{Summary and discussion}
\label{Discussion}

In this work we have proposed models for the quantum Rayleigh and \vdp\ oscillators. The mathematical and physical bases of the quantum Rayleigh and \vdp\ oscillators were discussed in Secs.~\ref{Quant_Ray+vdP} and \ref{PhysResults} respectively. In these sections we saw how certain aspects of the quantum models and their physics can be understood qualitatively from an understanding of their classical counterparts in Sec.~\ref{Class_Ray+vdP}. We then considered a driven oscillator in Sec.~\ref{Qsync} where frequency entrainment is demonstrated in the strongly-nonlinear regime. We also showed that relaxation oscillations are to be found in the mode of the Wigner function for the driven oscillator. Interestingly, relaxation oscillations in quantum mechanics can be effected by either a diffuse-and-zap mechanism, or a disappear-and-reappear mechanism in phase space depending on the strength of the oscillator's nonlinearity compared to the drive.

While our analyses and results on the quantum Rayleigh and \vdp\ oscillators are the first to go beyond the weakly-nonlinear regime, they are also still quite rudimentary. In forthcoming work we intend to study the single-oscillator physics in a more systematic fashion and investigate quantum synchronisation in coupled oscillators in the strongly-nonlinear regime. We expect these models to be useful for further exploration in quantum synchronisation and nonlinear dynamics as it now permits one to access an entirely new parameter regime in the quantum domain.

\begin{acknowledgments}

We thank Jofre Pedregosa Gutierrez for lending us his computing resources and expertise. We thank Dariel Mok for useful discussions and for pointing out minor errors in our manuscript. We would also like to thank Thi Ha Kyaw for some feedback on our manuscript. AC and LCK are supported by the Ministry of Education, Singapore and the National Research Foundation, Singapore. CN was supported by the National Research Foundation of Korea (NRF) grant funded by the Korea government (MSIT) (NRF-2019R1G1A1097074).

\end{acknowledgments}

\appendix

\section{Interpreting the master equation for the Rayleigh oscillator}
\label{InterpretRayleigh}

Recall that the amplitude equation defining the Rayleigh oscillator is
\begin{align}
\label{SMadotRectModel}
	\an{\ann}' = & - i \, \wo\, \an{\ann} + \frac{\mu}{2} \, ( q^2_0 - 1 ) \, \big[ \an{\ann} + \an{\adg} \big]  \nn \\
	                         & - \frac{\mu}{6} \, \big[ \an{\ann^3} + \an{\adg{}^3} \big]  - \frac{\mu}{2} \, \big[ \an{\adg \, \ann^2} + \an{\adg{}^2 \, \ann} \big]  \; .
\end{align}
We claimed that \eqref{SMadotRectModel} can be generated by a master equation of the form
\begin{align}
\label{LrayApp}
	\rho' = \Lcal_\sqr \, \rho = {}& - i \, \big[ \Hhat_\sqr\,, \rho \big] + \mu\,\big(q^2_0-1\big) \, \Dcal\big[ \adg \big] \rho  \nn  \\
	                                                        & + \frac{3\mu}{4} \; \Dcal\big[\ann^2\big] \rho + \mu \; \Dcal\big[\adg\ann - \half\,\adg{}^2\big] \rho   \; ,
\end{align}
where 
\begin{align}
\label{Hray}
	\hat{H}_\sqr = {}& \wo \, \adg\ann - i \, \eta_\sqr \, \big( \ann^2 - \adg{}^2 \big) + i \,\beta_\sqr \, \big( \ann^4 - \adg{}^4 \big)  \nn  \\
	                                 & + i \,\zeta_\sqr \, \big( \adg \ann^3 - \adg{}^3 \ann \big) \;,
\end{align}
with the coefficients $\eta_\sqr$, $\beta_\sqr$, and $\zeta_\sqr$ given by
\begin{align}
\label{Hsqrcoeff}
	\eta_\sqr = \mu\,(q^2_0-1)/4  \; ,  \quad   \beta_\sqr = \mu/24  \; ,    \quad   \zeta_\sqr = \mu/12  \; .
\end{align}

\subsection{Coherent processes}

The Hamiltonian terms in \eqref{LrayApp} can all be understood physically as degenerate parametric processes of different orders in nonlinear optics. The term proportional to $\ann^2 - \adg{}^2$ generates squeezing which is  realised using a nonlinear crystal with second-order susceptibility (degenerate parametric down conversion). The next two terms, proportional to $\ann^4 - \adg{}^4$ and $\adg \ann^3 - \adg{}^3 \ann$ correspond to processes mediated by a crystal of fourth-order nonlinear susceptibility. The Hamiltonian involving $\ann^4 - \adg{}^4$ is understood as a form of higher-order squeezing in quantum optics \cite{BM87,ZDH18} and can be realised as degenerate parametric down conversion with four photons. In principle it can also be implemented using a superconducting cavity, in which $\ann^3+\adg{}^3$ along with its non-degenerate variants have been implemented recently \cite{CSFD+20}. The next term, $\adg \ann^3 - \adg{}^3 \ann$, can be implemented using a similar method in quantum optics \cite{DADSI06}, and is no more difficult to realise than $\ann^4 - \adg{}^4$. In principle it can also be implemented using a superconducting cavity \cite{CSFD+20}. 
If we denote the set of external control parameters such as the nonlinear susceptibility and external fields by $\chi$, the Hamiltonian so obtained should then be written as
\begin{align}
\label{Hchi}
	\Hhat_\chi =  & - i \, \eta_\chi\, \big( \ann^2 - \adg{}^2 \big) + i \,\beta_\chi \, \big( \ann^4 - \adg{}^4 \big)  \nn  \\
	                          & + i \,\zeta_\chi \, \big( \adg \ann^3 - \adg{}^3 \ann \big)  \; .
\end{align}
For a given $\mu$ and $q_0$, we must then choose $\chi$ such that $\eta_\chi$, $\beta_\chi$, and $\zeta_\chi$ match \eqref{RayCoeff} of the main text in the absence of other processes. Realisations of such higher-order nonlinear parametric processes have been proposed in quantum optics using atoms and virtual photons \cite{KMMSN17}. 

\subsection{Incoherent processes}

\subsubsection{Linear amplification and nonlinear loss}

An ion-trap realisation for the first two dissipative channels in \eqref{LrayApp} already exists (see Ref.~\cite{LS13}). Here we simply comment further on alternative views of these terms. The dissipator $\kup\,\Dcal[\adg]$ can be understood to implement an ideal linear amplifier whose gain is specified by $\mu$ and $q_0$ \cite{Aga12,CHFKV19,CDGMS10}. One possible implementation is to use two-mode squeezing with the ancillary mode prepared in the vacuum state \cite{CCJP12}. The result is a parametric amplifier whose gain is now a function of the squeezing parameter. We can also realise $\kup\,\Dcal[\adg]$ using a gain medium in which all the atoms are maintained in the excited state with a strong pump \cite{Aga12}. In this case $\kup$ is the effective excited-state population of the gain medium. The next dissipative superoperator in \eqref{LrayApp}, $\kddn\,\Dcal[\ann^2]$, is simply a two-photon absorber \cite{Boy08,DH14} which can be realised using a nonlinear medium with third-order susceptibility and with all the atoms in the ground state \cite{Boy08,Bas95}. In this case $\kddn$ is the effective ground-state population of the nonlinear medium, specified by $\mu$. It is a well-studied process in quantum optics and is often referred to as a nonlinear absorber without qualification (see e.g.~Refs.~\cite{Rit90,GG93} and the references therein).

\subsubsection{Measurement and feedback}

The superoperator $\mu\,\Dcal[\adg\ann-\adg{}^2/2]$ is difficult to interpret. Unlike the other terms in the master equation discussed above, we do not know the process corresponding to $\Dcal[\adg\ann-\adg{}^2/2]$, and hence no way of knowing the meaning of its coefficient $\mu$. Here we overcome this by showing how this term can be realised with Markovian feedback \cite{WM10,Jac14,CW11a,CW11b}. That is, the system without feedback is given by
\begin{align}
\label{NoFBME}
	\rho' = \Lcal_{\chi}\, \rho = - i \big[\Hhat_\chi , \rho \big] + \kup \, \Dcal\big[\adg\big] \rho + \kddn \Dcal\big[\ann^2\big] \rho   \; ,
\end{align}
where $\Hhat_\chi$ is given by \eqref{Hchi} while the dissipative coefficients have to be tuned so that 
\begin{align}
	\kup = \mu \, (q^2_0-1) \; ,  \quad   \kddn = \frac{3\mu}{4}  \; . 
\end{align}
We have not written the dissipative coefficients in \eqref{NoFBME} directly in terms of $\mu$ to emphasise their dependence on external parameters. To use Markovian feedback one first measures an observable $\xhat$ of the system and conditions the system state on the measurement record continuously in time. Let us label the state so obtained as $\rho_{\rm c}(t)$. We then feed back the measurement outcomes (obtained with measurement strength $k$) by coupling it to a system observable $\hat{f}$ according to
\begin{align}
\label{Vfb}
	\Vhat_{\rm fb}(t) = \bigg[ \an{\xhat}_{\rm c}(t) + \frac{\xi(t)}{\rt{8k}} \bigg] \, \hat{f} \; ,
\end{align}
where $\an{\xhat}_{\rm c}(t)=\Tr[\xhat \rho_{\rm c}(t)]$. The measurement is noisy, modelled by the inclusion of a Gaussian white-noise process $\xi(t)$. It can then be shown that averaging over all possible conditioned states $\rho_{\rm c}(t)$ leads to the master equation 
\begin{align}
\label{FBME}
	\rho' = \Lcal_{\rm fb} \, \rho = - i\big[\Hhat_{\rm fb}, \rho \big] + 2 \, k \, \Dcal\big[\, \xhat - i\hat{f}/4k \,\big] \rho \; .
\end{align}
Note that as a consequence of performing Markovian feedback we also effect a Hamiltonian term, given by
\begin{align}
\label{Hfb}
	\Hhat_{\rm fb} = \frac{1}{4} \, \big( \xhat \, \hat{f} + \hat{f} \, \xhat \big)  \; .
\end{align}
It is simple to check that the desired dissipative channel can be achieved by the following choice of measurement and the feedback observable 
\begin{align}
\label{FBObservable}
	\xhat = \adg \ann - \frac{1}{4} \, \big( \ann^2 + \adg{}^2 \big)  \; ,   \quad    \hat{f} = i \, k \, \big( \ann^2 - \adg{}^2 \big)  \; . 
\end{align} 
From \eqref{FBME} and \eqref{Hfb} this leads to
\begin{align}
\label{LfbRay}
	\rho' = \Lcal_{\rm fb} \, \rho = - i\big[\Hhat_{\rm fb}, \rho \big] + 2\,k \, \Dcal\big[\, \adg\ann - \half \adg{}^2 \,\big] \rho   \; ,
\end{align}
where $\Hhat_{\rm fb}$ is now given by
\begin{align}
\label{HfbRay}
	\Hhat_{\rm fb} =  {}& i \, \frac{k}{2} \; \big( \ann^2 - \adg{}^2 \big) -  i \, \frac{k}{8} \; \big( \ann^4 - \adg{}^4 \big)  \nn \\
	                                       & + i \, \frac{k}{2} \; \big( \adg \ann^3 - \adg{}^3 \ann \big)  \; .
\end{align}
The full master equation with measurement and feedback is then 
\begin{align}
	\rho' = \big(  \Lcal_\chi + \Lcal_{\rm fb} \big) \rho =  {}& - i \big[ \Hhat_\sqr , \rho \big] +  \kup \, \Dcal\big[\adg\big] \rho + \kddn \Dcal\big[\ann^2\big] \rho  \nn \\
	                 & + 2\,k \, \Dcal\big[\, \adg\ann - \half\adg{}^2 \,\big] \rho  \; , 
\end{align}
where $\Hhat_\sqr$ has the form given in \eqref{Hray} but now the coefficients are modified by feedback, giving
\begin{align}
\label{HcoeffFB}
	\eta_\sqr = \eta_{\chi} - \frac{k}{2}  \; ,    \quad   \beta_\sqr =  \beta_\chi  - \frac{k}{8}  \; ,    \quad  \zeta_\sqr = \zeta_\chi + \frac{k}{2}  \; .
\end{align}
They are required to satisfy \eqref{Hsqrcoeff}. Given a desired value for $\mu$, we should measure $\xhat$ at a strength of $k=\mu/2$ which gives $2k=\mu$.

This now gives us a way to interpret $\mu\,\Dcal[\adg\ann-\adg{}^2/2]$. First, the physical meaning of its coefficient is now clear---it is the measurement strength in a measurement of $\xhat$. Second, the process associated with $\Dcal[\adg\ann-\adg{}^2/2]$ is one of measurement and feedback. In particular, the form of $\hat{f}$ in \eqref{FBObservable} means the feedback defined by \eqref{Vfb} is implemented by squeezing the system by an amount proportional to the measurement outcome. Third, it is possible to put $\xhat$ as defined in \eqref{FBObservable} in a form that makes it more amenable to interpretation. To do so we note that 
\begin{align}
\label{DTrans}
	\mu \, \Dcal\big[\adg\ann-\half\adg{}^2 \big]  \; \stackrel{\an{\ann}}{=} \; \frac{\mu}{2} \, \Dcal\big[\,\adg\ann-\adg{}^2\big] + \frac{3\mu}{2} \, \Dcal\big[\ann\big]   \; ,
\end{align}
where $\stackrel{\an{\ann}}{=}$ means that \eqref{SMadotRectModel} remains unchanged on using \eqref{DTrans}. The effect of the extra linear dissipative channel arising from this transformation is to add more noise to the oscillator. It is the simplest form of dissipation in bosonic systems, and typically occurs when the system is in contact with an environment, even for a zero-temperature environment. If we were to think of the various degenerate parametric processes along with the linear gain, and nonlinear loss as implemented through crystals and gain media, (i.e.~atom-photon interactions) in a cavity system, then the extra linear dissipator would simply model energy leakage through a cavity mirror at some prescribed rate determined by $\mu$. In this case we would replace the no-feedback master equation \eqref{NoFBME} by 
\begin{align}
\label{NoFBMEDTrans}
	\rho' = {}& \Lcal_{\chi}\, \rho  \\
                     = {}& - i \big[\Hhat_\chi , \rho \big] + \kup \, \Dcal\big[\adg\big] \rho + \kddn \Dcal\big[\ann^2\big] \rho + \kdn \, \Dcal[\ann] \rho    \nn  \; ,
\end{align}
where everything is as before except that $\kdn=3\mu/2$. By following the same measurement and feedback protocol, we can show that $\mu\,\Dcal\big[\,\adg\ann-\adg{}^2\big]/2$ can be implemented by choosing the following measurement and feedback observables with $k=\mu/4$,
\begin{gather}
\label{FBObservableAlt}
	\xhat = \adg \ann - \frac{1}{2} \, \big( \ann^2 + \adg{}^2 \big) = \frac{:\phat^2:}{2} \; ,   \\
          \hat{f} = i \, 2\,k \, \big( \ann^2 - \adg{}^2 \big)  \; . 
\end{gather}
Recall that we have defined $\phat=-i(\ann-\adg)$ and $:\!h(\ann,\adg)\!:$ denotes normal ordering of the $\ann$ and $\adg$ appearing in the function $h$. This form shows that $\xhat$ corresponds to something physical, being the quadrature autocorrelation function at zero-delay for a zero-mean field. However, as we already highlighted in the main text in Sec.~\ref{GeneralRemarks}, the in-loop measurement here would need to be a direct one as opposed to the usual method of homodyning.

\section{Interpreting the master equation for the \vdp\ oscillator}
\label{vdPApp}

The same procedure using measurement and feedback works for the \vdp\ oscillator as well just as it did above. However, if we interchange the forms of $\xhat$ and $\hat{f}$ with slight modifications, it becomes possible to implement $\Dcal[\adg\ann-\adg{}^2/2]$ as well as the Hamiltonian $i\,\zeta_\dia(\adg\ann^3-\adg{}^3\ann)$ in the \vdp\ model. Thus our measurement and feedback scheme is more suited to the \vdp\ oscillator. We follow the same feedback protocol as above, except that this time, we choose the observables for measurement and feedback to be
\begin{gather}
	\xhat = - \frac{i}{4} \, \big( \,\ann^2 - \adg{}^2 \big)  = \frac{:\qhat\phat:}{4}\, ,   \\
       \hat{f} = 4 \, k \, \adg \ann -  k \, \big( \ann^2 + \adg{}^2\big)   \; . 
\end{gather}
A simple calculation then shows that this leads to \eqref{LfbRay} with 
\begin{align}
	2 \, k \, \Dcal\big[\, \xhat - i\hat{f}/4k \,\big] = {}& 2 \, k \, \Dcal\big[\!-\!i\big(\adg\ann-\half\adg{}^2\big)\big]  \\
	                                                                                         = {}& 2 \, k \, \Dcal\big[\, \adg\ann-\half\adg{}^2 \,\big]  \; ,
\end{align}
and 
\begin{align}
	\Hhat_{\rm fb} = {}& - i \, \frac{k}{2} \big( \ann^2 - \adg{}^2 \big) - i \, \frac{k}{2} \big( \adg \ann^3 - \adg{}^3 \ann \big)  \nn \\
	                                      & + i \, \frac{k}{8} \, \big( \ann^4 - \adg{}^4 \big)  \; .
\end{align}
It is clear that choosing $k=\mu/2$ as before allows us to generate $i\,\zeta_\dia(\adg\ann^3-\adg{}^3\ann)$ and $\mu\,\Dcal[\adg\ann-\adg{}^2/2]$. The remaining terms can in principle be realised as discussed in the previous section.

\section{Fidelity of the driven Rayleigh oscillator}
\label{FidApp}

\begin{figure}[t]
\centerline{\includegraphics[width=0.46\textwidth]{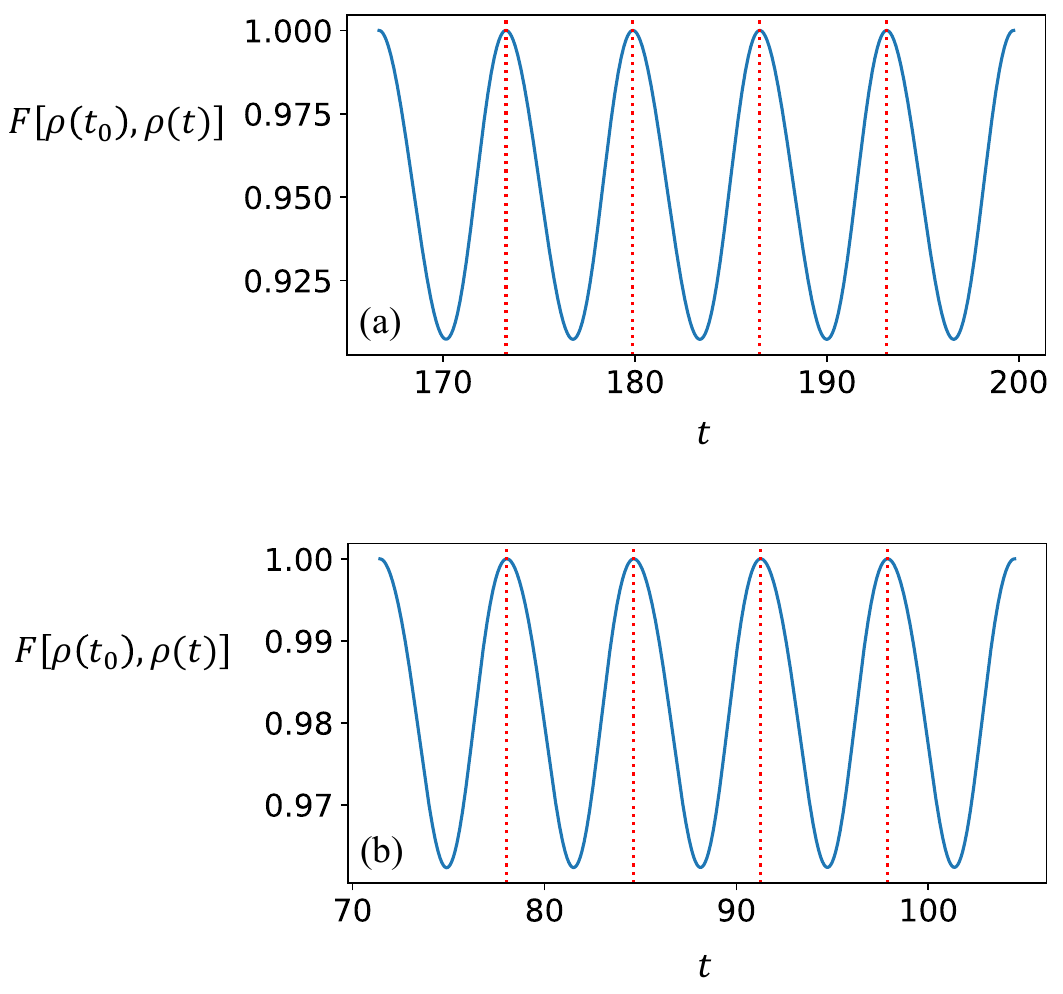}}
\caption{\label{Fidelity} Fidelities between $\rho(t_0)$ and $\rho(t)$ for $q_0=2$, $\wo=1$, $\omega_1=0.95$, $\phi=\pi/2$. Vertical red dotted lines show $t$ values at $t_0+nT$ for $n=1,2,3,4$. (a) $\mu=0.03$,  $\epsilon=0.1$, and $t_0=5/\mu=166.67$,   (b) $\mu=0.07$, $\epsilon=0.1$, and $t_0=5/\mu=71.43$. }
\end{figure}

Here we show the driven Rayleigh oscillator of Sec.~\ref{Qsync} has a state whose long-time limit is periodic with a period equal to the drive's. Recall that our model is given by
\begin{align}
\label{RayApp}
	\rho' = -i \, \epsilon \,\cos(\omega_1 \, t + \phi) \, \big[ \ann+\adg, \rho \big] + \Lcal_\sqr \, \rho  \; .
\end{align}
\begin{figure}[t]
\centerline{\includegraphics[width=0.38\textwidth]{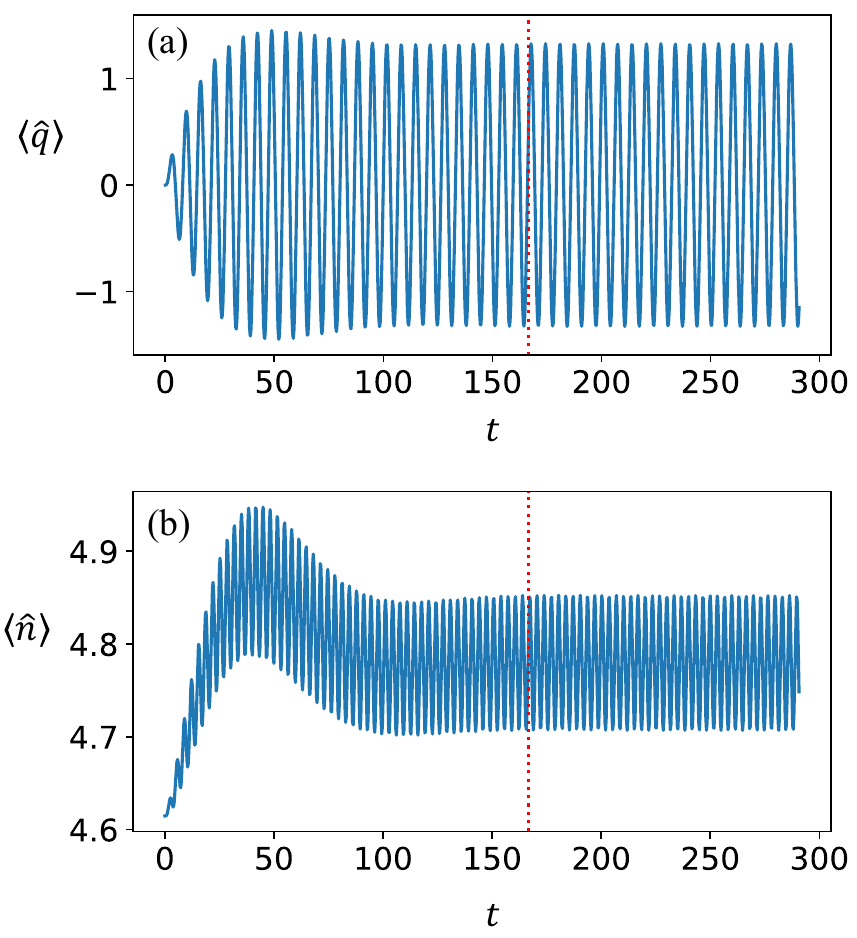}}
\caption{\label{AppCtmaxSmallmu} Dynamics of the driven Rayleigh oscillator starting from the undriven steady state. $\mu=0.03$, $q_0=2$, $\wo=1$, $\omega_1=0.95$, $\phi=\pi/2$, and $\epsilon=0.1$. Red dotted line marks $t=5/\mu$.}
\end{figure}
The period of the drive is thus given by $T=2\pi/\omega_1$. If $\rho(t)$ is periodic in the long-time limit, then the fidelity between it and a fixed state $\rho(t_0)$ must also be periodic for $t>t_0$ when $t_0 \longrightarrow \infty$. The fidelity between $\rho(t_0)$ and $\rho(t)$ is given by 
\begin{align}
\label{F(rho(t0),rho(t))}
	F\big[ \rho(t_0), \rho(t) \big] = \Tr \Big\{ \Big[ \rt{\rho(t_0)} \rho(t) \rt{\rho(t_0)} \Big]^{\frac{1}{2}} \Big\}  \; .
\end{align}
We will therefore plot $F\big[ \rho(t_0), \rho(t) \big]$ as a function of $t$ for a fixed but large $t_0$. This is shown in Fig.~\ref{Fidelity} for the same parameter values as those used in Fig.~\ref{SyncSmallmu}\,(a) and Fig.~\ref{SyncLargemu}(a) of the main text (repeated again in the fidelity figure captions here for convenience).
\begin{figure}[b]
\centerline{\includegraphics[width=0.38\textwidth]{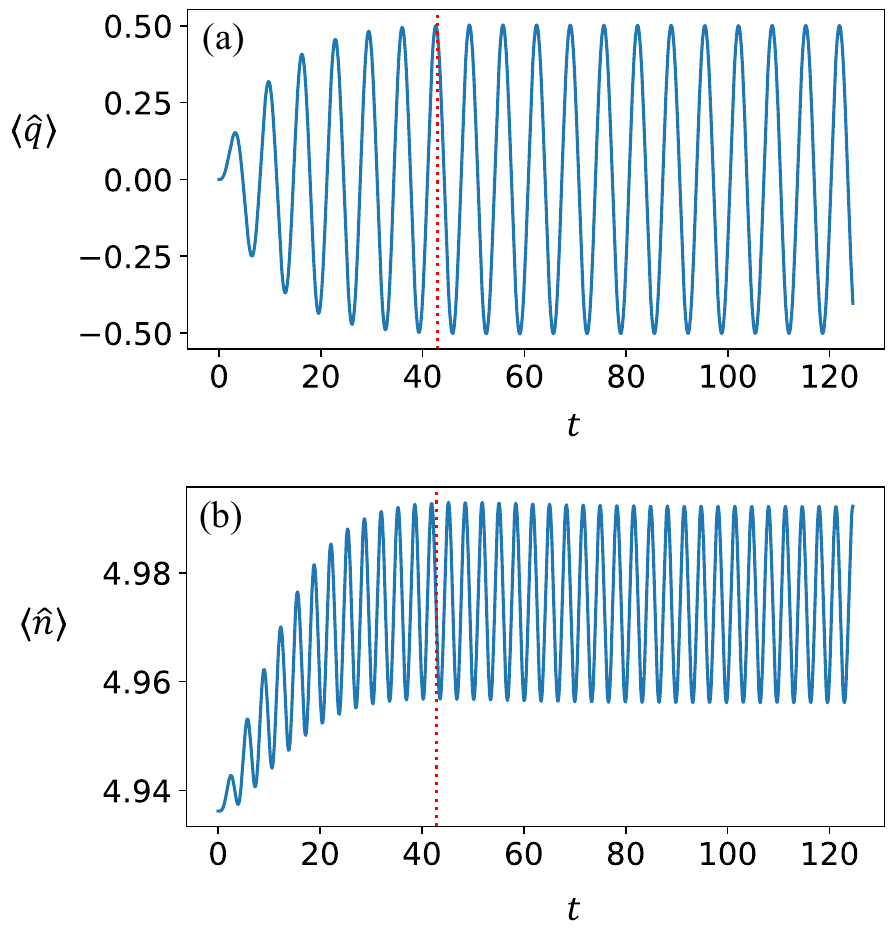}}
\caption{\label{AppCtmaxLargemu} Dynamics of the driven Rayleigh oscillator starting from the undriven steady state. $\epsilon=0.06$, $\mu=0.07$. Red dotted line marks $t=3/\mu$.}
\end{figure}

In Fig.~\ref{Fidelity} we have plotted \eqref{F(rho(t0),rho(t))} from $t_0$ to $t_0+5T$ (five periods of the drive). Note that in practice $t_0$ just needs to be sufficiently large so that all transient effects are avoided. We have marked integer multiples of the drive's period with vertical red dotted lines. As can be seen, the fidelity starts at one as it should for $t=t_0$, and subsequently returns to one at integer multiples of the drive's period. To make sure that our $t_0$ values suffice we show explicitly the transient dynamics of the driven oscillator for the average position and average number of quanta in Fig.~\ref{AppCtmaxSmallmu}\,(a) and (b) respectively. The time $5/\mu$ [$t_0$ value in Fig.~\ref{Fidelity}\,(a) and Figs.~\ref{RelaxOscUnimodal} and \ref{QuantumSlimeUni} of the main text] is marked with a vertical red dotted line and is clearly well beyond the transient dynamics. The same method can be applied for Fig.~\ref{Fidelity}\,(b), but instead we will illustrate the transient dynamics corresponding to Figs.~\ref{RelaxOscBimodal} and \ref{QuantumSlimeBi} of the main text for variety. Again, the average position and number of quanta for this case is shown in Fig.~\ref{AppCtmaxLargemu}\,(a) and (b). The time $3/\mu$ ($t_0$ value for Figs.~\ref{RelaxOscBimodal} and \ref{QuantumSlimeBi}) is marked with a vertical red dotted line. Here we can see that it is approximately the onset of the long-time limit.

\section{Stationarity of the driven Rayleigh oscillator}
\label{CorrFuncApp}
\begin{figure}[t]
\centerline{\includegraphics[width=0.43\textwidth]{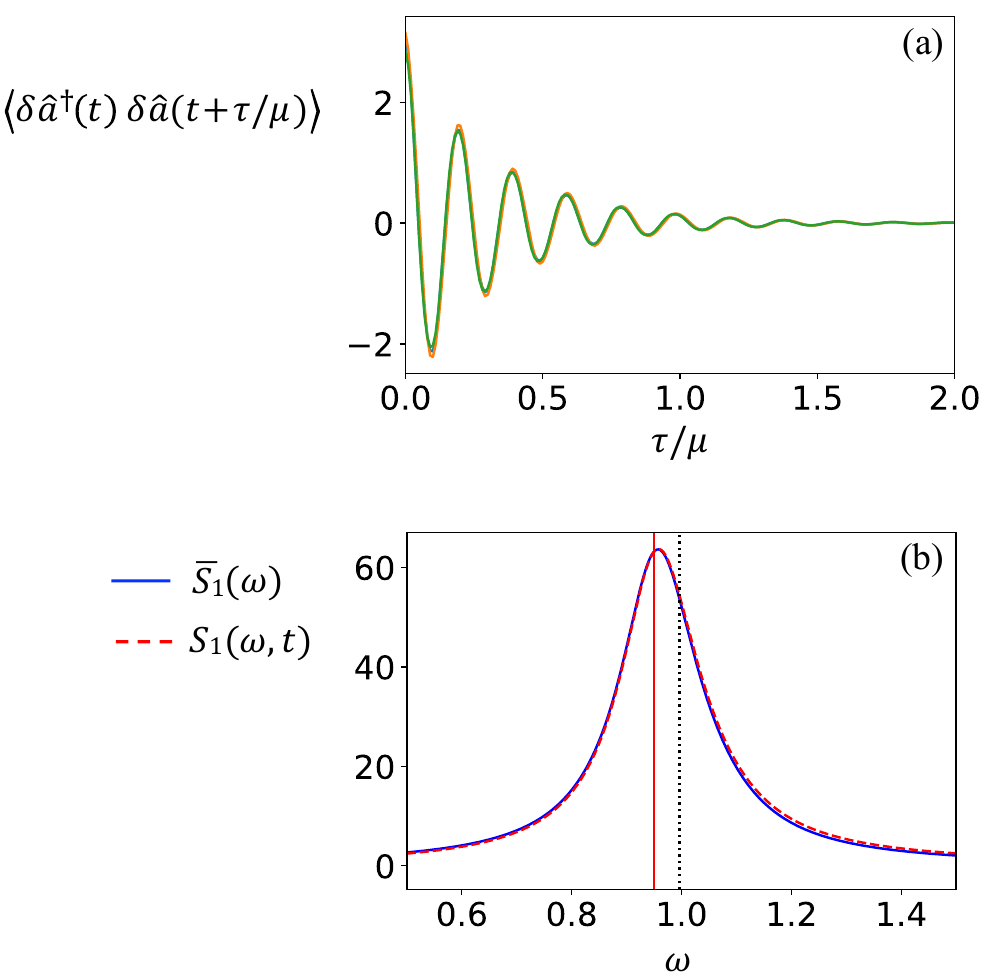}}
\caption{\label{CorrFunc3} Approximate second-order stationarity of $\delta\ann(t)$ for $\mu=0.03$, $\omega_0 = 1$, $q_0 = 2$, and $\omega_0=1$, $\omega_1 = 0.95$, and $\epsilon=0.5$. (a) Correlation functions corresponding to three evenly spaced values of $t$ in the interval $[t_0,t_0+T]$ for $t_0=5/\mu$. (b) Comparison between $\overline{S}(\omega)$ averaged over three different $t$ values (blue solid), and $S_1(\omega,t)$ for one value of $t$ (red dashed line). The vertical red solid line is $\omega_1$ while the vertical black dotted line is $\Omega_0$.}
\end{figure}

We mentioned in Sec.~\ref{Entrainment} that the power spectrum of the driven Rayleigh oscillator is nearly independent of the time origin. Recall from \eqref{S(w,t)} that the power spectrum is given by
\begin{align}
\label{S1(w,t)App}
	S_1(\omega,t) = \int^\infty_{-\infty} d\tau \; e^{i\omega \tau} \, \ban{\delta\adg(t) \, \delta\ann(t+\tau)}  \; .
\end{align}
To illustrate the weak dependence of \eqref{S1(w,t)App} on $t$ we consider the six parameter sets for $(\mu,\epsilon)$: $(0.03, 0.1)$, $(0.03, 0.3)$, $(0.03, 0.5)$, $(0.07,  0.1)$, $(0.07, 0.3)$, and $(0.07, 0.5)$. Among these six parameter sets, $(0.03, 0.5)$ show the most dependence on $t$.

In Fig.~\ref{CorrFunc3}\,(a) we plot three different correlation functions, corresponding to three different values of $t$ against $\tau/\mu$. From this we can see that the different correlation functions are nearly overlapping one another. The average of these correlation functions is then computed and used to calculate $\overline{S}_1(\omega)$. We then compare $\overline{S}_1(\omega)$ to $\overline{S}_1(\omega,t)$ for a single value of $t$. This is shown in Fig.~\ref{CorrFunc3} where the two spectra can be seen to nearly overlap as one might expect from seeing Fig.~\ref{CorrFunc3}\,(a). We have in fact calculated correlation functions corresponding to 20 different values of the time origin (not shown) and have found them to be approximately overlapping as seen in Fig.~\ref{CorrFunc3}.

\end{document}